\documentclass[epsfig,12pt]{article}
\usepackage{makeidx}
\usepackage{amsmath}
\usepackage{amsfonts}
\usepackage{amssymb}
\usepackage{graphicx}

\input epsf.sty
\newtheorem{theorem}{Theorem}
\newtheorem{acknowledgement}[theorem]{Acknowledgement}

\makeatletter \@addtoreset{equation}{section}
\renewcommand{\theequation}{\thesection.\arabic{equation}}
\input{tcilatex}
\begin{document}

\title{\vspace{-2cm}%
\rightline{\mbox{\small
{LPHE-MS-1501-re}} \vspace {1cm}} \textbf{\ }$\mathcal{N}=2$ \textbf{\
Supersymmetry Partial Breaking and Tadpole Anomaly}}
\author{R.Ahl Laamara$^{1,2}$, M. N. El Kinani$^{1}$, E. H Saidi$^{1,2}$, M.
Vall$^{1,2}$ \\
{\small 1. Lab Of High Energy Physics, Modeling and Simulations, }\\
{\small Faculty Of Sciences, Rabat, Morocco}\\
{\small 2. Centre of Physics and Mathematics, CPM- Morocco}}
\maketitle

\begin{abstract}
We consider the\textbf{\ }$U\left( 1\right) ^{n}$ extension of the effective 
$\mathcal{N}=2$ supersymmetric\textbf{\ }$U\left( 1\right) \times U\left(
1\right) $\textbf{\ }model of \textrm{arXiv:1204.2141}; and study the
explicit relationship between partial breaking of $\mathcal{N}=2$
supersymmetry constraint and D3 brane tadpole anomaly of type IIB string on
Calabi-Yau threefolds in presence of H$^{RR}$ and H$^{NS}$ fluxes. We also
comment on supersymmetry breaking in the particular $\mathcal{N}=2$\textbf{\ 
}$U\left( 1\right) $ Maxwell theory; and study its interpretation in
connection with the tadpole anomaly with extra localized flux sources.

\end{abstract}


\section{Introduction}

Breaking $\mathcal{N}=2$ supersymmetric quantum field theories in 4d space
time at two different mass scales has been subject of interest for many
years \textrm{\cite{1A}-\cite{1AE}; and refs therein}. This scenario is
possible in 4d $\mathcal{N}=2$ supergravity theory; but not with $\mathcal{N}%
=2$ global supersymmetry suspected to live at lower energies below Planck
scale. If gravity is neglected, superalgebra relations requires that once
one of the two \emph{global} supercharges $Q_{\alpha }^{\pm }=(Q_{\alpha },%
\tilde{Q}_{\alpha })$ is broken; say $Q_{\alpha }^{-}$, the second $%
Q_{\alpha }^{+}$ has to be broken too. However this constraint can be
bypassed in the presence of magnetic Fayet-Iliopoulos (FI) terms induced by
non perturbative BPS states such as D- branes of type II strings. With non
zero magnetic FI couplings, the supercurrent algebra develops a constant
term that violates the SU$_{R}\left( 2\right) $ R-symmetry of the
supercharges \textrm{\cite{6A}}; offering as a consequence a way to break $%
\mathcal{N}=2$ supersymmetry partially via gauginos instead of gravitinos 
\textrm{\cite{4A,1B,2B,3B,4B}}. This idea has been approached in past by
using non linear realisation of half of the eight supersymmetric charges 
\textrm{\cite{1C,2C,3C,4C,5C,6C,7C}}; but further developed recently in 
\textrm{\cite{1D,2D,3D}} by using $\mathcal{N}=1$ superspace QFT$_{4}$
method where a simple, but instructive, effective $\mathcal{N}=2$
supersymmetric abelian \textrm{U}$\left( 1\right) ^{2}$ model, with two
breaking scales $\Lambda _{1}$ and $\Lambda _{2}$, has been engineered.

\  \  \  \  \newline
In this paper, we consider the partial breaking of $\mathcal{N}=2$
supersymmetry in the effective 4d $U\left( 1\right) \times U\left( 1\right) $
model of \textrm{\cite{1D,2D}}, to which we refer now on as ADJ effective
gauge theory; and study explicitly its relationship with D3 brane tadpole
anomaly of type IIB string compactified on local Calabi-Yau threefolds
(CY3).\ To deal with brane realisation of the ADJ construction, we first
relax the rank of the abelian group symmetry by considering the effective $%
U\left( 1\right) ^{2}$ model as the leading prototype in the family of 4d $%
\mathcal{N}=2$ $U\left( 1\right) ^{n}$ gauge models indexed by $n\geq 2$;
and then think of this set of abelian gauge models in terms of an effective
low energy theory following from D3 branes wrapping 3-cycles in type IIB
string on local CY3 with a n- dimensional symplectic homology basis of
3-cycles $\left( A^{a},B_{a}\right) $, a=1,...,n. In this D3 brane
realisation of 4d $\mathcal{N}=2$ $U\left( 1\right) ^{n}$ gauge theory,
partial breaking of global $\mathcal{N}=2$ supersymmetry is induced by H$%
_{3}^{RR}$ and H$_{3}^{NS}$ fluxes; and the ADJ condition $\sum_{a}\frac{%
g_{a}}{\kappa _{a}}=0$ supporting the partial breakings is interpreted in
terms of conservation of the total 3-forms flux $\Phi _{flux}$ in the
Calabi-Yau threefolds; that is $\Phi _{flux}=\int_{CY3}H_{3}^{NS}\wedge
H_{3}^{RR}=0$. We also study the missing $n=1$ term in the sequence of 4d $%
\mathcal{N}=2$ $U\left( 1\right) ^{n}$ gauge models with $n\geq 2$; this
particular model, which corresponds type IIB string on conifold geometry, is
anomalous in agreement with known results in literature; this anomaly may be
directly learnt from the naive extension of ADJ condition which is given by
the singular equation $\frac{g}{\boldsymbol{\kappa }}=0$ requiring the
vanishing of the gauge coupling constant $g=0$ for finite magnetic FI
coupling $\kappa $. By trying to engineer a 4d $\mathcal{N}=2$ $U\left(
1\right) $ model going beyond ADJ constraint by\ deforming the singularity
like $\nu +\frac{g}{\boldsymbol{\kappa }}=0$, we end with an explicit
breaking of $\mathcal{N}=2$ supersymmetry down to $\mathcal{N}=1$. The brane
interpretation of this deformation in terms of presence of D7 branes or O3
planes is also studied by using a result from \textrm{\cite{1G}}.

\  \  \  \  \newline
The presentation is as follows: In section 2, we review the basis of the
effective 4d $\mathcal{N}=2$ $U\left( 1\right) ^{2}$ model; and derive the $%
\mathcal{N}=2$ ADJ constraint equation and its $\mathcal{N}=1$ deformation.
In section 3, we give the main lines of the $\mathcal{N}=1$ superfield
formulation of the 4d $\mathcal{N}=2$ $U\left( 1\right) ^{n}$ gauge theory
describing the gauge dynamics of n $\mathcal{N}=2$ gauge multiplets coupled
to a single tensor multiplet. In section 4, we study the realisation of ADJ
model in type IIB string on CY3 with non trivial fluxes of the 3-form field
strengths H$_{3}^{NS}$ and H$_{3}^{RR}$.\newline
In section 5, we give conclusion and make some comments. In section 6 we
give two appendices; in the first appendix we collect some useful tools on
type IIB string compactification to 4d space time; and in the second we
describe the gauge and supersymmetric transformations of single-tensor and
Maxwell multiplets in superspace.

\section{$\mathcal{N}=2$ $U\left( 1\right) ^{n}$ theory and ADJ constraints}

In this section, we review the main lines of the model of \textrm{\cite%
{1D,2D}; and study particular aspects of ADJ constraint equation \
supporting the partial supersymmetry breaking in this theory. }Following 
\textrm{\cite{1D,2D,3D}, the ADJ model }is a 4d $\mathcal{N}=2$
supersymmetric effective gauge theory where gravity is decoupled; but global 
$\mathcal{N}=2$ supersymmetry is broken at two different scales. The
simplest version of the model realising the two partial breaking is given by
the interacting dynamics of a $\mathcal{N}=2$ single tensor multiplet $%
\mathcal{T}^{\left( \mathcal{N}=2\right) }$ with two abelian $\mathcal{N}=2$
gauge supermultiplets $\mathcal{V}_{1}^{\left( \mathcal{N}=2\right) }$ and $%
\mathcal{V}_{2}^{\left( \mathcal{N}=2\right) }$. Besides special Kahler
geometry of the gauge background, the model has a Chern-Simons type
interaction between $\mathcal{T}^{\left( \mathcal{N}=2\right) }$ and the
linear combination $g_{1}\boldsymbol{V}_{1}^{\left( \mathcal{N}=2\right)
}+g_{2}\boldsymbol{V}_{2}^{\left( \mathcal{N}=2\right) }$.

\subsection{ADJ theory in $\mathcal{N}=1$ superspace}

Because of lack of a simple formulation of 4d supersymmetric gauge theories
with \emph{8} supercharges in $\mathcal{N}=2$ superspace, one is limited to
use the standard $\mathcal{N}=1$ superspace method with the price that only
half of supersymmetries is manifestly exhibited; the other half is hidden;
but can be linearly realised in absence of magnetic FI couplings.

\subsubsection{Fibering $\mathcal{N}=2$ chiral superspace}

A way to deal with the \emph{4} hidden supersymmetric charges is to use $%
\mathcal{N}=2$ chiral superspace and think about it in terms of fibration of
two copies of $\mathcal{N}=1$ chiral superspaces; a $\mathcal{N}%
_{fiber}=1^{\prime }$ chiral superspace, with odd coordinates $\tilde{\theta}%
^{\alpha }$, fibered on a $\mathcal{N}_{base}=1$ chiral superspace base with
odd coordinates $\theta ^{\alpha }$. Schematically, this fibration may be
represented like 
\begin{equation}
\begin{tabular}{lll}
$\mathcal{N}_{fiber}=1^{\prime }$ & $\  \  \  \rightarrow $ \  \  \  & $\mathcal{N%
}=2$ \\ 
&  & $\  \left. 
\begin{array}{c}
\text{ }\downarrow \\ 
\end{array}%
\right. $ \\ 
&  & $\mathcal{N}_{base}=1$%
\end{tabular}
\label{ZIP}
\end{equation}%
\begin{equation*}
\end{equation*}%
In this chiral superspace fibration, typical $\mathcal{N}=2$ chiral
superfields have the structure $\boldsymbol{\Phi }^{\mathcal{N}=2}=%
\boldsymbol{\Phi }(z,\theta ,\tilde{\theta})$ with space time coordinate $%
z^{\mu }$ related to the real $x^{\mu }$ by two pure imaginary shifts $%
i\upsilon ^{\mu }+i\tilde{\upsilon}^{\mu }$, one from $\mathcal{N}%
_{fiber}=1^{\prime }$ fiber and the other from the $\mathcal{N}_{base}=1$
base as shown on the relation $z=y-i\tilde{\theta}\sigma \widetilde{\bar{%
\theta}}$ with $y=x-i\theta \sigma \bar{\theta}$. Viewed from fiber, $%
\boldsymbol{\Phi }^{\mathcal{N}=2}$ can be expanded in a finite series of $%
\tilde{\theta}$ as follows 
\begin{equation}
\boldsymbol{\Phi }^{\mathcal{N}=2}=\Phi ^{\mathcal{N}=1}+\sqrt{2}\text{ }%
\tilde{\theta}^{\alpha }\Psi _{\alpha }^{\mathcal{N}=1}+\tilde{\theta}^{2}F^{%
\mathcal{N}=1}+...  \label{1}
\end{equation}%
with expansion modes given by $\mathcal{N}_{base}=1$ superfields: $\Phi ^{%
\mathcal{N}=1}=\Phi \left( y,\theta \right) $\ and similarly for $\Psi
_{\alpha }^{\mathcal{N}=1}$ and $F^{\mathcal{N}=1}$. The extra dots stand
for additional terms involving space time derivatives generated by $-i\tilde{%
\theta}\sigma ^{\mu }\widetilde{\bar{\theta}}\partial _{\mu }=-i\tilde{%
\upsilon}^{\mu }\partial _{\mu }$. The expansion modes in (\ref{1}) describe 
$\mathcal{N}=1$ chiral superfields in the base; and are related amongst
others by those $\mathcal{N}_{fiber}=1^{\prime }$ supersymmetric
transformations in the fiber; for example 
\begin{eqnarray}
\tilde{\delta}\Phi ^{\mathcal{N}=1} &=&\sqrt{2}\tilde{\varepsilon}^{\alpha
}\Psi _{\alpha }^{\mathcal{N}=1}  \notag \\
\tilde{\delta}\Psi _{\alpha }^{\mathcal{N}=1} &=&\sqrt{2}\tilde{\varepsilon}%
^{\alpha }F^{\mathcal{N}=1}-\frac{i\sqrt{2}}{2}\sigma ^{\mu }\widetilde{\bar{%
\varepsilon}}\partial _{\mu }\Phi ^{\mathcal{N}=1} \\
\tilde{\delta}F^{\mathcal{N}=1} &=&-\frac{i\sqrt{2}}{2}\text{ }\partial
_{\mu }\Psi _{\alpha }^{\mathcal{N}=1}\sigma ^{\mu }\widetilde{\bar{%
\varepsilon}}  \notag
\end{eqnarray}%
By imposing appropriate constraint relations on $\boldsymbol{\Phi }^{%
\mathcal{N}=2}$, one obtains the desired $\mathcal{N}_{base}=1$ superfields
to describe supersymmetric matter or radiation with \emph{8} supercharges.
In this way of doing, $\mathcal{N}=2$ supersymmetric gauge multiplet is then
approached by using superfield strength $\mathcal{W}^{\mathcal{N}=2}$ with
expansion as in (\ref{1}); but satisfying moreover $D^{\alpha }\tilde{D}%
_{\alpha }\mathcal{W}^{\mathcal{N}=2}+hc=0$. As this constraint involves
both the chiral $\mathcal{W}^{\mathcal{N}=2}$ and its adjoint conjugate, one
ends with a $\tilde{\theta}$- expansion involving both $\mathcal{N}=1$
chiral $\boldsymbol{X}$ and $\bar{D}^{2}\boldsymbol{\bar{X}}_{{}}$ as follows%
\begin{equation}
\begin{tabular}{llll}
$\mathcal{W}^{\mathcal{N}=2}$ & $=$ & $\boldsymbol{X}+i\sqrt{2}\text{ }%
\tilde{\theta}^{\alpha }\boldsymbol{W}_{\alpha }-\tilde{\theta}^{2}\left( 
\frac{1}{4}\bar{D}^{2}\boldsymbol{\bar{X}}_{{}}\right) $ &  \\ 
$\mathcal{\tilde{W}}^{\mathcal{N}=2}$ & $=$ & $\mathcal{W}^{\mathcal{N}=2}+%
\tilde{\theta}^{2}\frac{1}{2\kappa }$ & 
\end{tabular}
\label{22}
\end{equation}%
where the role of the extra constant coefficient $\frac{1}{2\kappa }$ will
be discussed later on; it scales as $mass^{2}$ seen that $\left[ \mathcal{W}%
^{\mathcal{N}=2}\right] =\left[ \boldsymbol{X}\right] =mass^{1}$ and the $%
\mathcal{N}_{base}=1$ chiral gauge superfield strength spinor $\left[ 
\boldsymbol{W}_{\alpha }\right] =mass^{3/2}$; it may be generated by the
particular and \emph{asymmetric} shift of the $\tilde{\theta}^{2}$ component%
\begin{equation}
\left. \frac{1}{4}\bar{D}^{2}\boldsymbol{\bar{X}}_{{}}\right \vert _{\theta
=0}=\bar{F}^{\bar{X}}\qquad \rightarrow \qquad \bar{F}^{\bar{X}}-\frac{1}{%
2\kappa }  \label{m}
\end{equation}%
By asymmetric we mean eq(\ref{m}) but without modifying the $\boldsymbol{X}$
superfield in (\ref{22}). This property may be roughly interpreted as giving
a non zero VEV to the $\tilde{\theta}^{2}$- component field of expansion of (%
\ref{22}) as 
\begin{equation}
\left \langle \frac{1}{4}\bar{D}^{2}\boldsymbol{\bar{X}}_{{}}\right \rangle =%
\frac{1}{\sqrt{2}}\left \langle A+iB\right \rangle =-\frac{1}{2\kappa }
\label{kp}
\end{equation}%
breaking thus the $\mathcal{N}_{fiber}=1^{\prime }$ supersymmetry in fiber.
For a stringy interpretation of the coupling constant $\frac{1}{\kappa }$ in
terms of 3- form flux through non compact 3-cycles in local CY3; see eq(\ref%
{kap}). \newline
A quite similar expansion is valid for the $\mathcal{N}=2$ tensor multiplet $%
\mathcal{T}^{\left( \mathcal{N}=2\right) }$ which is described as well by a
constrained $\mathcal{N}=2$ chiral superfield \textrm{\cite{1D,2D,3D}, see
also appendix }$\mathrm{\S 6.2}$; it is given by%
\begin{equation}
\mathcal{T}^{\mathcal{N}=2}=\boldsymbol{Y}+\sqrt{2}\text{ }\tilde{\theta}%
^{\alpha }\boldsymbol{\chi }_{\alpha }-\tilde{\theta}^{2}\left( \frac{i}{2}%
\boldsymbol{\Phi }+\frac{1}{4}\bar{D}^{2}\boldsymbol{\bar{Y}}_{{}}\right)
\label{T}
\end{equation}%
with $\mathcal{N}_{fiber}=1^{\prime }$ supersymmetric transformations in
fiber as%
\begin{equation}
\begin{tabular}{lll}
$\tilde{\delta}_{{}}\boldsymbol{Y}$ & $\boldsymbol{=}$ & $+\sqrt{2}\text{ }%
\tilde{\epsilon}^{\alpha }\boldsymbol{\chi }_{\alpha }$ \\ 
$\tilde{\delta}_{{}}\boldsymbol{\chi }_{\alpha }$ & $=$ & $\sqrt{2}\tilde{%
\epsilon}_{\alpha }\mathcal{E}-\frac{i}{\sqrt{2}}\sigma _{\alpha \dot{\alpha}%
}^{\mu }\widetilde{\bar{\epsilon}}^{\dot{\alpha}}\partial _{\mu }\boldsymbol{%
Y}$ \\ 
$\tilde{\delta}_{{}}\mathcal{E}$ & $=$ & $\frac{\sqrt{2}}{2i}\partial _{\mu }%
\boldsymbol{\chi }_{\alpha }\sigma _{\alpha \dot{\alpha}}^{\mu }\widetilde{%
\bar{\epsilon}}^{\dot{\alpha}}$%
\end{tabular}
\label{TR}
\end{equation}%
where we have set%
\begin{equation}
\mathcal{E}=-\frac{i}{2}\boldsymbol{\Phi }-\frac{1}{4}\bar{D}^{2}\boldsymbol{%
\bar{Y}}_{{}}
\end{equation}%
With these tools at hand, we turn to describe useful features on superfield
spectrum of $\mathcal{N}=2$ supersymmetric ADJ model. For later use, we will
give both the $\mathcal{N}=2$ chiral superfields spectrum and their
splitting in terms of $\mathcal{N}_{base}=1$ superfields.

\  \  \  \ 

\emph{\ more on matter sector}\newline
The matter sector of ADJ model is quite simple; it involves one 4d $\mathcal{%
N}=2$ matter multiplet having two dual realisations as given by eqs(\ref{A1}%
). \newline
In the realisation we will be using in present study, $\mathcal{N}=2$ matter
is described by a $\mathcal{N}=2$ chiral superfield $\mathcal{T}^{\left( 
\mathcal{N}=2\right) }$ with expansion along fiber direction as in (\ref{T}%
). From the $\mathcal{N}_{base}=1$ base view, this expansion has four chiral
superfields: two bosonic $\boldsymbol{Y}$, $\boldsymbol{\Phi };$ and a
fermionic superfield doublet $\boldsymbol{\chi }_{\alpha }=\left( 
\boldsymbol{\chi }_{1},\boldsymbol{\chi }_{2}\right) $; altogether they
capture $16+16$ off shell degrees of freedom. This number may be reduced
down to $8+8$ by thinking about $\boldsymbol{Y}$ as an exotic auxiliary
superfield playing the role of a Lagrange superfield parameter capturing the
constraint on partial breaking of second supersymmetry; and about $%
\boldsymbol{\chi }_{\alpha }$ as a superfield prepotential of a hermitian
linear multiplet $\boldsymbol{L}$ given by the relation 
\begin{equation}
\boldsymbol{L}=D^{\alpha }\boldsymbol{\chi }_{\alpha }+\bar{D}_{\dot{\alpha}%
}^{{}}\boldsymbol{\bar{\chi}}_{{}}^{\dot{\alpha}}  \label{L}
\end{equation}%
Observe that $\boldsymbol{L}$ is invariant under the change $\boldsymbol{%
\chi }_{\alpha }^{\prime }=\boldsymbol{\chi }_{\alpha }+\frac{i}{4}\bar{D}%
\boldsymbol{^{2}}D_{\alpha }\Omega $ with $\Omega $ an arbitrary real
superfield; this symmetry together with footnote$^{1}$ allows to reduce the $%
16+16$ degrees of freedom down to $8+8$; for details see \textrm{appendix }$%
\mathrm{\S 6.2}$; other features can be found in \textrm{\cite{1D,2D,3D}}.%
\newline
For completeness, notice that using a result on hypermultiplet duality on
superfield representations of $\mathcal{N}=2$ matter multiplet \textrm{\cite%
{1E,2E,3E,4E,5E}}, we can show that $\mathcal{N}=2$ superfield $\mathcal{T}%
^{\left( \mathcal{N}=2\right) }$ has two dual representations in terms of $%
\mathcal{N}_{base}=1$ superfields; the $\left( \boldsymbol{\Phi },%
\boldsymbol{L}\right) $ we will be using in this paper; and a second
realisation based on two chiral superfields $\boldsymbol{Q}_{1},$ $%
\boldsymbol{Q}_{2}$;%
\begin{equation}
\begin{tabular}{lllllll}
$a)$ & : & $\mathcal{T}^{\left( \mathcal{N}=2\right) }$ & $\equiv $ & $%
\boldsymbol{\Phi },$ $\boldsymbol{L}$ & $;$ & $\boldsymbol{Y}$ \\ 
$b)$ & : & $\mathcal{T}^{\left( \mathcal{N}=2\right) }$ & $\equiv $ & $%
\boldsymbol{Q}_{1},$ $\boldsymbol{Q}_{2}$ & $;$ & $\boldsymbol{Y}$%
\end{tabular}
\label{A1}
\end{equation}%
\  \  \ 

\emph{ADJ constraint}\newline
First notice that in above (\ref{A1}), it looks like if we have three $%
\mathcal{N}_{base}=1$ superfields to describe $\mathcal{N}=2$ matter; this
is not exact since $\boldsymbol{Y}$ is some how a "spurious superfield"
carrying no physical degrees of freedom; it is a topological object
exhibiting very special properties as shown by eqs(\ref{Y}-\ref{y}); this is
our reason behind putting $\boldsymbol{Y}$ aside in eq(\ref{A1}); it breaks $%
\mathcal{N}=2$ supersymmetry partially and is one of the nice observations
in \textrm{\cite{2D}}; there it appears as a Lagrange superfield parameter
capturing a constraint relation $f\left( g_{a},\kappa _{a}\right) =0$ of the
model, which to fix ideas may be thought of as 
\begin{equation}
f\left( g_{a},\kappa _{a}\right) =\sum_{a=1}^{n}\frac{g_{a}}{\kappa _{a}}%
=0\qquad ,\qquad n\geq 2  \label{cst}
\end{equation}%
giving a relationship between the coupling constants $g_{a}$ and the
magnetic FI couplings $1/\kappa _{a}$ of the ADJ model; see also eq(\ref{rc}%
) given below for explicit details. In the limit 
\begin{equation}
\frac{1}{\kappa _{a}}\rightarrow 0
\end{equation}%
the above constraint is trivially solved and then $\boldsymbol{Y}$ has no
role to play in this $\mathcal{N}=2$ supersymmetric limit.\newline
Notice also that the sum on integer n in eq(\ref{cst}) rules out the
particular case $n=1$; since the corresponding condition reads as 
\begin{equation}
\frac{g_{1}}{\kappa _{1}}=0
\end{equation}%
leading to $g_{1}=0$ for finite $1/\kappa _{1}$; and then no ADJ theory with
one U$\left( 1\right) $ gauge factor \textrm{\cite{6C,7C,1D,2D,3D}}. By
trying to overcome the constraint $\frac{g_{1}}{\kappa _{1}}=0$ by adding an
extra term like 
\begin{equation}
\nu +\frac{g_{1}}{\kappa _{1}}=0  \label{stc}
\end{equation}%
with $\nu $ a real parameter having same scaling mass dimension as $\frac{%
g_{a}}{\kappa _{a}}$; one breaks \emph{explicitly} $\mathcal{N}=2$
supersymmetry down to $\mathcal{N}=1$.

\subsubsection{$\mathcal{N}=1$ superfields in $U\left( 1\right) ^{2}$ ADJ
model}

In our superspace description of $U\left( 1\right) \times U\left( 1\right) $
ADJ model, we will use a particular set of $\mathcal{N}=1$ superfields;
these are the chiral $\boldsymbol{\Phi }$ and hermitian $\boldsymbol{L}$ for
representing $\mathcal{T}^{\left( \mathcal{N}=2\right) }$; and 2 gauge
superfields $\left( \boldsymbol{V}_{1},\boldsymbol{V}_{2}\right) $, 2 chiral 
$\left( \boldsymbol{X}_{1},\boldsymbol{X}_{2}\right) $ for representing $%
\mathcal{V}_{1}^{\left( \mathcal{N}=2\right) }\oplus \mathcal{V}_{2}^{\left( 
\mathcal{N}=2\right) }$. Let us comment this system of superfields.

\  \  \  \  \  \  \  \  \  \  \  \ 

$\bullet $ $\mathcal{T}^{\left( \mathcal{N}=2\right) }$ \emph{sector} 
\newline
In the $\mathcal{N}=1$ superfield realisation given by the first relation of
eqs(\ref{A1}), the dynamics of $\mathcal{T}^{\left( \mathcal{N}=2\right) }$
is described by two basic superfields and an auxiliary one; these are:

\begin{description}
\item[$\left( i\right) $] the chiral superfield $\boldsymbol{\Phi }$ with
the usual $\theta $-expansion namely a leading scalar component $\phi $; a
Weyl fermions $\psi _{\alpha }$ and auxiliary field F$_{\phi }$;

\item[$\left( ii\right) $] the standard hermitian linear multiplet $%
\boldsymbol{L}$ satisfying the superspace constraint relations $D^{2}%
\boldsymbol{L}=\bar{D}\boldsymbol{^{2}L=0}$ following from (\ref{L}); this
is a particular superfield with $\theta $- expansion in component fields as
follows%
\begin{equation}
\begin{tabular}{lll}
$\boldsymbol{L}$ & $=$ & $C+i\theta .\eta -i\bar{\theta}.\bar{\eta}+\theta
\sigma ^{\mu }\bar{\theta}\varepsilon _{\mu \nu \rho \sigma }\partial ^{\nu
}B^{\rho \sigma }+$ \\ 
&  & $\frac{1}{2}\theta ^{2}\bar{\theta}\bar{\sigma}^{\mu }\partial _{\mu
}\eta -\frac{1}{2}\bar{\theta}^{2}\theta \sigma ^{\mu }\partial _{\mu }\bar{%
\eta}-\frac{1}{4}\theta ^{2}\bar{\theta}^{2}\square C$%
\end{tabular}
\label{216}
\end{equation}%
involving the propagating real field $C$ and the field strength of the
antisymmetric tensor field $B^{\rho \sigma }$; but no auxiliary field. The
superfields $\boldsymbol{\Phi }$ and $\boldsymbol{L}$ are related under
fiber $\mathcal{N}_{fiber}=1^{\prime }$ supersymmetric variations as follows 
\begin{eqnarray}
\tilde{\delta}_{\epsilon }\boldsymbol{L} &=&\frac{\sqrt{2}}{2i}\left( \tilde{%
\epsilon}_{{}}^{\alpha }D_{\alpha }\boldsymbol{\Phi -}\widetilde{\bar{%
\epsilon}}_{\dot{\alpha}}\bar{D}_{{}}^{\dot{\alpha}}\boldsymbol{\bar{\Phi}}%
\right)  \notag \\
\tilde{\delta}_{\epsilon }\boldsymbol{\Phi } &=&-i\sqrt{2}\widetilde{\bar{%
\epsilon}}_{\dot{\alpha}}\bar{D}_{{}}^{\dot{\alpha}}\boldsymbol{L}
\label{29}
\end{eqnarray}%
with%
\begin{equation}
\left[ \tilde{\delta}_{\epsilon ^{\prime }},\tilde{\delta}_{\epsilon }\right]
\boldsymbol{\Psi }=-2i\left( \tilde{\epsilon}_{{}}\sigma ^{\mu }\widetilde{%
\bar{\epsilon}}^{\prime }-\tilde{\epsilon}_{{}}^{\prime }\sigma ^{\mu }%
\widetilde{\bar{\epsilon}}_{{}}\right) \partial _{\mu }\boldsymbol{\Psi }
\end{equation}%
with $\boldsymbol{\Psi }$ standing for $\boldsymbol{L}$ and $\boldsymbol{%
\Phi }$.

\item[$\left( iii\right) $] an extra auxiliary superfield $\boldsymbol{Y}$
capturing information on \textrm{non linear realisation} of the $\mathcal{N}%
_{fiber}=1^{\prime }$ hidden supersymmetry; it is not needed for the closure
of transformations (\ref{29}); but will be used to approach partial
supersymmetry breaking.\ Properties of this superfield have been explored in 
\textrm{\cite{1D,2D}} where, using gauge fixing\textrm{\footnote{%
Following analysis of appendix 8.2, the reduction of the $16+16$ degrees of $%
\mathcal{N}=2$ chiral superfield $\mathcal{T}^{N=2}$ down to $8+8$ is
achieved into two steps: a first reduction from $16+16$ down to $12+12$
ensured by gauge symmetry under $\boldsymbol{\chi }_{\alpha }^{\prime }=%
\boldsymbol{\chi }_{\alpha }+\frac{i}{4}\bar{D}^{2}D_{\alpha }\Omega $ (\ref%
{gh}); a second reduction from $12+12$ down to $8+8$ given by requiring
symmetry under gauge transformation $Y\rightarrow Y-\frac{1}{2}\bar{D}%
^{2}\Upsilon $ with $\Upsilon $ a real superfield. Gauge fixing of this
symmetry leads precisely to eq(\ref{Y}); for more details see also \cite{1D}.%
}} method, it has been shown to have the following remarkable $\theta $-
expansion%
\begin{equation}
\boldsymbol{Y}^{gauged}=\frac{i}{4!}\theta ^{2}\varepsilon ^{\mu \nu \rho
\sigma }C_{\mu \nu \rho \sigma }  \label{Y}
\end{equation}%
for details see appendix 6.2. This relation shows that $\boldsymbol{Y}$
encodes data on the constant antisymmetric tensor $C_{\mu \nu \rho \sigma
}=4!\varepsilon _{\mu \nu \rho \sigma }\Delta $. Because of its special
dependence in $\theta $, $\boldsymbol{Y}$ has no physical degrees of freedom%
\begin{equation}
\delta _{\epsilon }\boldsymbol{Y}^{gauged}\boldsymbol{=0}
\end{equation}%
and obeys moreover a nilpotency property 
\begin{equation}
\boldsymbol{Y}^{\dagger }\boldsymbol{Y}=\theta ^{4}\Delta ^{2}\qquad ,\qquad 
\boldsymbol{YY}=0=\boldsymbol{Y}^{\dagger }\boldsymbol{Y}^{\dagger }
\label{y}
\end{equation}
\end{description}

\  \  \  \newline
In the $\mathcal{N}=1$ superfield realisation given by the second relation
of eqs(\ref{A1}), the role of $\boldsymbol{\Phi }$ and $\boldsymbol{L}$ gets
played by the two chiral superfields $\boldsymbol{Q}_{1}$ and $\boldsymbol{Q}%
_{2}$ capturing opposite charge under a $U\left( 1\right) $ gauge symmetry.
The duality transformations between the two matter multiplet realisations
are given by Legendre transform in superspace \textrm{\cite{1E,2E,3E,4E,5E}}%
; they may be written as follows%
\begin{equation}
\begin{tabular}{lll}
$\boldsymbol{Q}_{1}$ & $=$ & $2^{\frac{-1}{4}}\sqrt{\boldsymbol{\Phi }}e^{+%
\boldsymbol{\Phi }^{\prime }}$ \\ 
$\boldsymbol{Q}_{2}$ & $=$ & $2^{\frac{-1}{4}}\sqrt{\boldsymbol{\Phi }}e^{-%
\boldsymbol{\Phi }^{\prime }}$%
\end{tabular}
\label{dt}
\end{equation}%
with $\boldsymbol{\Phi }$ as in the first relation of eqs(\ref{A1}) and
where $\boldsymbol{\Phi }^{\prime }$ is another chiral superfield. We will
not need this realisation in this paper; but to fix ideas we give some
comments on their dynamics in $\S $ 3.2; see eq(\ref{hp}).

\  \  \  \ 

$\bullet $ $\mathcal{V}_{a}^{\left( \mathcal{N}=2\right) }$\emph{gauge sector%
}\newline
The gauge sector of ADJ supersymmetric U$\left( 1\right) ^{2}$ model
involves two $\mathcal{N}=2$ abelian Maxwell type multiplets described by
the hermitian superfields $\boldsymbol{V}_{1}^{\left( \mathcal{N}=2\right) }$
and $\boldsymbol{V}_{2}^{\left( \mathcal{N}=2\right) }$ with superfields
strength $\theta $- expansions along fiber direction as in eq(\ref{1}).
Following \textrm{\cite{1D,2D}}, the solution of constraint equations lead
to the $\mathcal{N}=1$ superfields spectrum 
\begin{equation}
\begin{tabular}{lllll}
$\boldsymbol{V}_{1}^{\left( \mathcal{N}=2\right) }$ & $\equiv $ & $%
\boldsymbol{V}_{1},$ $\boldsymbol{X}_{1}$ & , & $\boldsymbol{\kappa }_{1}$
\\ 
$\boldsymbol{V}_{2}^{\left( \mathcal{N}=2\right) }$ & $\equiv $ & $%
\boldsymbol{V}_{2},$ $\boldsymbol{X}_{2}$ & , & $\boldsymbol{\kappa }_{2}$%
\end{tabular}%
\end{equation}%
where the hermitian $\boldsymbol{V}_{1},$ $\boldsymbol{V}_{2}$ are the usual 
$\mathcal{N}_{base}=1$ gauge superfield potentials; and where $\boldsymbol{X}%
_{1},$ $\boldsymbol{X}_{2}$ are two chiral superfields. So the gauge
symmetry of the model is $U_{1}\left( 1\right) \times U_{2}\left( 1\right) $%
. The extra $\boldsymbol{\kappa }_{1},$ $\boldsymbol{\kappa }_{2}$ are
constants and are as in (\ref{22}); they may be put in correspondence with
the auxiliary chiral superfield $\boldsymbol{Y}$ as it may be viewed by
comparing (\ref{T}) with (\ref{22}); that is:%
\begin{equation}
\sum \frac{g_{a}}{\boldsymbol{\kappa }_{a}}\qquad \leftrightarrow \qquad 
\boldsymbol{Y}
\end{equation}%
The general form of the superspace lagrangian density $\mathcal{L}_{gauge}$
of the gauge superfields depending on the prepotential $\mathcal{F}\left( 
\boldsymbol{X}_{1},\boldsymbol{X}_{2}\right) $ reads as follows%
\begin{equation}
\mathcal{L}_{gauge}=\mathcal{L}_{_{_{U_{1}\left( 1\right) \times U_{2}\left(
1\right) }}}+\mathcal{L}_{_{FI}}  \label{ag}
\end{equation}%
with gauge lagrangian density in $\mathcal{N}_{base}=1$ superspace given by%
\begin{equation}
\mathcal{L}_{_{_{U_{1}\left( 1\right) \times U_{2}\left( 1\right) }}}=\int
d^{4}\theta _{{}}\frac{i}{2}\left( \mathcal{\bar{F}}_{a}\boldsymbol{X}^{a}-%
\mathcal{F}_{a}\boldsymbol{\bar{X}}^{a}\right) +\int d^{2}\theta _{{}}\left(
-\frac{i}{4}\mathcal{F}_{ab}W^{a}.W^{b}\right) +hc  \label{ga}
\end{equation}%
and Fayet-Iliopoulos part as%
\begin{equation}
\mathcal{L}_{_{FI}}=\int d^{4}\theta _{{}}\xi _{a}\boldsymbol{V}^{a}-\int
d^{2}\theta _{{}}\frac{e_{a}}{4}\boldsymbol{X}^{a}-\int d^{2}\theta _{{}}%
\frac{i}{4\kappa _{a}}\mathcal{F}_{a}+hc  \label{AG}
\end{equation}%
where the real $\xi _{a}$ and complex \textrm{(pure imaginary)} $e_{a}$ are
constants and where the holomorphic $\mathcal{F}_{a}=\frac{\partial \mathcal{%
F}}{\partial X^{a}}$ and $\mathcal{F}_{ab}=\frac{\partial ^{2}\mathcal{F}}{%
\partial X^{a}\partial X^{b}}.$\ 

\subsection{ADJ constraint and $\mathcal{N}=1$ deformation}

Here we use the $\mathcal{N}=1$ superfield spectrum of ADJ model to study
the derivation of the constraint eq(\ref{cst}) and its $\mathcal{N}=1$
deformation (\ref{stc}).

\subsubsection{Superfield $\boldsymbol{Y}$}

Viewed from $\mathcal{N}=2$ chiral superspace, the $U\left( 1\right) \times
U\left( 1\right) $ ADJ supersymmetric model involves the $\mathcal{N}=2$
chiral superfields $\mathcal{W}_{a}^{\mathcal{N}=2}$ given by (\ref{22});
and the $\mathcal{T}^{\mathcal{N}=2}$ of (\ref{T}). These $\mathcal{N}=2$
chiral superfields are remarkable; they have the same scaling mass dimension
and quite similar $\tilde{\theta}$- expansions which make them to share some
general features. Indeed, though physically different objects, the
resemblance between their $\tilde{\theta}$- expansions could serve as a
guide to have more insight into the ADJ construction. This formal property
has been used in \textrm{\cite{1D,2C}} to study the interaction between a $%
\mathcal{N}=2$ Maxwell multiplet $\mathcal{W}^{\mathcal{N}=2}$ and a tensor $%
\mathcal{T}^{\mathcal{N}=2}$. There, the formal similarity between the two
chiral superfields $\mathcal{W}^{\mathcal{N}=2}$ and $\mathcal{T}^{\mathcal{N%
}=2}$; in particular their scaling mass dimension and $\theta $- expansions,
has been used to build the linear combination of these $\mathcal{N}=2$
chiral superfield 
\begin{equation}
\mathcal{W}^{\mathcal{N}=2}+2g\mathcal{T}^{\mathcal{N}=2}  \label{wt}
\end{equation}%
to reach the gauge invariant quantity%
\begin{equation}
\mathcal{F}_{\mu \nu }^{Max}-gB_{\mu \nu }
\end{equation}%
that plays a central role in the $\mathcal{N}=2$ Dirac-Born-Infeld U$_{\max
}\left( 1\right) $ theory; and also in studying electric-magnetic duality in 
$\mathcal{N}=2$ chiral superspace in presence of Chern-Simons coupling. In
this relation, $\mathcal{F}_{\mu \nu }^{Max}$ is the usual field strength of
the Maxwell gauge field potential; and $B_{\mu \nu }$ the antisymmetric
gauge potential appearing in the tensor multiplet. \  \  \  \newline
By exhibiting this formal similarity between $\mathcal{W}^{\mathcal{N}=2}$
and $\mathcal{T}^{\mathcal{N}=2}$; one finds that there exist a
correspondence between their $\mathcal{N}=1$ superfields contents; by
comparing the $\tilde{\theta}$- expansions (\ref{22}) and (\ref{T}); as well
as topological relations reported in appendix B (\ref{X2}) and (\ref{X3}),
one ends with%
\begin{equation}
\begin{tabular}{ll||ll}
\  \ gauge multiplet $\mathcal{\tilde{W}}^{\mathcal{N}=2}$ &  &  & tensor
multiplet $\mathcal{T}^{\mathcal{N}=2}$ \  \  \\ \hline
$\  \  \  \  \  \  \  \  \  \  \  \boldsymbol{X}$ &  &  & $\  \  \  \  \  \  \  \  \boldsymbol{Y%
}$ \\ 
$\  \  \  \  \  \  \  \  \  \  \ i\boldsymbol{W}_{\alpha }$ &  &  & $\  \  \  \  \  \  \  \ 
\boldsymbol{\chi }_{\alpha }$ \\ 
$\  \  \  \  \  \  \  \  \  \  \  \frac{-1}{2\kappa }$ &  &  & $\  \  \  \  \  \  \  \  \frac{i%
}{2}\boldsymbol{\Phi }$ \\ 
$\  \  \  \  \  \  \  \  \  \  \  \bar{D}^{2}\boldsymbol{\bar{X}}_{{}}$ &  &  & $\  \  \
\  \  \  \  \  \bar{D}^{2}\boldsymbol{\bar{Y}}_{{}}$ \\ \hline
\end{tabular}
\label{cor}
\end{equation}%
\begin{equation*}
\end{equation*}%
where $\boldsymbol{Y}$ occupies a place in $\mathcal{T}^{\mathcal{N}=2}$
that is similar to the place occupied by $\boldsymbol{X}$ in $\mathcal{%
\tilde{W}}^{\mathcal{N}=2}$. Obviously the superfields in left and right of
table (\ref{cor}) have different meanings and carry different degrees of
freedom; but as far as fibration of $\mathcal{N}=2$ supersymmetry is
concerned; this correspondence may be used as an indication to get more
insight into the general form of constraint equation captured by $%
\boldsymbol{Y}$.\ 

\subsubsection{Deriving ADJ condition}

The ADJ constraint equation is obtained by from $\mathcal{N}=2$ Chern-Simons
couplings between the linear combination of the gauge superfield strengths $%
(\sum_{a}g_{a}\mathcal{\tilde{W}}_{a}^{\mathcal{N}=2})$ and the tensor
multiplet $\mathcal{T}^{\mathcal{N}=2}$. In $\mathcal{N}=2$ chiral
superspace where $\mathcal{N}=2$ supersymmetry is manifest, this CS coupling
reads in terms of $\mathcal{\tilde{W}}_{a}^{\mathcal{N}=2}$ and $\mathcal{T}%
^{\mathcal{N}=2}$ as follows 
\begin{equation}
\mathcal{L}_{CS}=-2i\dint d^{2}\theta _{{}}d^{2}\tilde{\theta}_{{}}\left(
\dsum \limits_{a=1}^{n}g_{a}\mathcal{\tilde{W}}_{a}^{\mathcal{N}=2}\right) 
\mathcal{T}^{\mathcal{N}=2}
\end{equation}%
where $n=2$ for the case of U$\left( 1\right) ^{2}$ model; but can generally
take any value $n\geq 2$ as the case of U$\left( 1\right) ^{n}$ models with
n gauge $\mathcal{\tilde{W}}_{a}^{\mathcal{N}=2}$ coupled to $\mathcal{T}^{%
\mathcal{N}=2}$. By using (\ref{22}) and (\ref{T}) and performing
integration with respect to $\tilde{\theta}$; one brings above CS coupling
to the form%
\begin{equation}
\begin{tabular}{lll}
$\mathcal{L}_{CS}$ & $=$ & $+2\dint d^{4}\theta _{{}}\left( \dsum
\limits_{a=1}^{n}g_{a}\boldsymbol{V}^{a}\right) \boldsymbol{L}-\dint
d^{2}\theta _{{}}\left( \dsum \limits_{a=1}^{n}g_{a}\boldsymbol{X}%
^{a}\right) \boldsymbol{\Phi }$ \\ 
&  & $-i\dint d^{2}\theta _{{}}\left( \dsum \limits_{a=1}^{n}\frac{g_{a}}{%
\kappa ^{a}}\right) \boldsymbol{Y}$%
\end{tabular}
\label{hc}
\end{equation}%
where $\mathcal{N}=1$ supersymmetry in the base of fibration is manifest and
the fibered $\mathcal{N}=1^{\prime }$ one becomes hidden. Because of linear
dependence, the superfield equation of $\boldsymbol{Y}$ leads to the
constraint%
\begin{equation}
\dsum \limits_{a=1}^{n}\frac{g_{a}}{\kappa ^{a}}=0\qquad ,\qquad n\geq 2
\end{equation}%
Notice that the deformation of the CS coupling (\ref{hc}) by adding the term 
$\mathrm{\nu }\int d^{2}\theta _{{}}\boldsymbol{Y}$ like%
\begin{eqnarray}
\mathcal{L}_{CS}^{\prime } &=&+2\dint d^{4}\theta _{{}}\left( \dsum
\limits_{a=1}^{n}g_{a}\boldsymbol{V}^{a}\right) \boldsymbol{L}-\dint
d^{2}\theta _{{}}\left( \dsum \limits_{a=1}^{n}g_{a}\boldsymbol{X}%
^{a}\right) \boldsymbol{\Phi }  \notag \\
&&-i\dint d^{2}\theta _{{}}\left( \mathrm{\nu }_{{}}+\dsum \limits_{a=1}^{n}%
\frac{g_{a}}{\kappa ^{a}}\right) \boldsymbol{Y}  \label{ob}
\end{eqnarray}%
preserves gauge symmetry as shown by (\ref{sg}) of appendix B; but breaks
explicitly $\mathcal{N}=2$ supersymmetry down to $\mathcal{N}=1$. Under this
deformation, the ADJ\ constraint becomes%
\begin{equation}
\mathrm{\nu }_{{}}+\dsum \limits_{a=1}^{n}\frac{g_{a}}{\kappa ^{a}}=0
\label{def}
\end{equation}

\section{Superspace Lagrangian}

The $\mathcal{N}=1$ superspace expression of the lagrangian density $%
\mathcal{L}$ describing the interacting dynamics of the above $\mathcal{N}=2$
supersymmetric system $\left \{ \mathcal{W}_{a}^{\left( \mathcal{N}=2\right)
},\mathcal{T}^{\left( \mathcal{N}=2\right) }\right \} $ can be approached in
two manners depending on the $\mathcal{N}=1$ superfield realisation used to
represent the $\mathcal{T}^{\left( \mathcal{N}=2\right) }$ single tensor
multiplet.

\subsection{Using $\mathcal{N}=1$ multiplets $\left( \boldsymbol{L},%
\boldsymbol{\Phi }\right) $}

With the realisation of the tensor multiplet $\mathcal{T}^{\left( \mathcal{N}%
=2\right) }$ in terms of the superfields the chiral $\boldsymbol{\Phi },$
the hermitian $\boldsymbol{L}$ as well as the Lagrange chiral superfield $%
\boldsymbol{Y}$ carrying the ADJ constraint; and following \textrm{\cite{2D}}%
, the superspace lagrangian density of the $\mathcal{N}=2$ supersymmetric U$%
\left( 1\right) ^{2}$ model describing coupled dynamics of $\mathcal{W}%
_{a}^{\left( \mathcal{N}=2\right) }$ and $\mathcal{T}^{\left( \mathcal{N}%
=2\right) }$ reads in $\mathcal{N}=1$ superspace as follows%
\begin{equation}
\mathcal{L}=\mathcal{L}_{gauge}+\mathcal{L}_{ST}+\mathcal{L}_{CS}  \label{TS}
\end{equation}%
with $\mathcal{L}_{gauge}$ as in eq(\ref{ga}) and 
\begin{eqnarray}
\mathcal{L}_{ST}+\mathcal{L}_{CS} &=&\int d^{4}\theta _{{}}\sqrt{\boldsymbol{%
L}^{2}+2\boldsymbol{\bar{\Phi}}_{{}}\boldsymbol{\Phi }}-\boldsymbol{L}\ln
\left( \boldsymbol{L+}\sqrt{\boldsymbol{L}^{2}+2\boldsymbol{\bar{\Phi}}_{{}}%
\boldsymbol{\Phi }}\right)  \notag \\
&&+2\int d^{4}\theta _{{}}g_{a}\boldsymbol{V}^{a}\boldsymbol{L}-\int
d^{2}\theta _{{}}\left( m+g_{a}\boldsymbol{X}^{a}\right) \boldsymbol{\Phi }
\label{ST} \\
&&-i\int d^{2}\theta _{{}}\left( \frac{g_{1}}{\kappa _{1}}+\frac{g_{2}}{%
\kappa _{2}}\right) \boldsymbol{Y}+hc  \notag
\end{eqnarray}%
In what follows, we first make few comments useful for our later analysis;
then we study scalar potential of the gauge model.\ 

\subsubsection{Properties of (\protect \ref{ST})}

From the above expression of the superspace lagrangian density (\ref{ST}),
we learn a set of special properties on the superfield realisation of matter
using single tensor multiplet; in particular the following:\newline
First, the mass constant $m$ in (\ref{ST}) can be absorbed by shifting the
linear combination $g_{a}\boldsymbol{X}^{a}$; it will be dropped out in what
follows. \newline
Second, the superfields $\boldsymbol{\Phi }$ and $\boldsymbol{L}$ are gauge
invariant and scale as $mass^{2}$; their coupling to the gauge multiplet is
of Chern-Simons type%
\begin{equation}
2\int d^{4}\theta _{{}}\left( g_{1}\boldsymbol{V}_{1}+g_{2}\boldsymbol{V}%
_{2}\right) \boldsymbol{L}-\int d^{2}\theta _{{}}\left( g_{1}\boldsymbol{X}%
_{1}+g_{2}\boldsymbol{X}_{2}\right) \boldsymbol{\Phi }
\end{equation}%
they involve the remarkable linear combinations $g_{1}\boldsymbol{V}%
_{1}+g_{2}\boldsymbol{V}_{2}$ and $g_{1}\boldsymbol{X}_{1}+g_{2}\boldsymbol{X%
}_{2}$. Moreover, the contribution of the superfield $\boldsymbol{Y}$ in the
full superspace lagrangian density (\ref{ST}) appears linearly as follows 
\begin{equation}
i\int d^{2}\theta _{{}}\left( \frac{g_{1}}{\kappa _{1}}+\frac{g_{2}}{\kappa
_{2}}\right) \boldsymbol{Y}+hc
\end{equation}%
together with the particular linear combination $\frac{g_{1}}{\kappa _{1}}+%
\frac{g_{2}}{\kappa _{2}}$. So the auxiliary superfield $\boldsymbol{Y}$ in
ADJ theory plays the role of a Lagrange superfield capturing the constraint
relation 
\begin{equation}
\frac{g_{1}}{\kappa _{1}}+\frac{g_{2}}{\kappa _{2}}=0  \label{rc}
\end{equation}%
showing that the ratio $\frac{g_{1}}{g_{2}}$ of the two gauge couplings is
fixed by the ratio $\frac{\kappa _{1}}{\kappa _{2}}$ of the magnetic FI
coupling constants. \newline
Third, the kinetic energy density of $\boldsymbol{\Phi }$ and $\boldsymbol{L}
$ involves non polynomial expressions, a square root term $\sqrt{\boldsymbol{%
L}^{2}+2\boldsymbol{\bar{\Phi}}_{{}}\boldsymbol{\Phi }}$ and a logarithm one
namely $\boldsymbol{L}\ln \left( \boldsymbol{L+}\sqrt{\boldsymbol{L}^{2}+2%
\boldsymbol{\bar{\Phi}}_{{}}\boldsymbol{\Phi }}\right) $; this non linearity
may be understood as due to the antisymmetric field $B_{\mu \nu }$. Self
interactions of $\left( \boldsymbol{\Phi },\boldsymbol{L}\right) $ are also
non polynomial and are generally characterized by an arbitrary hermitian
prepotential $H\left( \boldsymbol{\Phi ,\bar{\Phi}};\boldsymbol{L}\right) $
with superspace lagrangian density as \textrm{\cite{1E,2E,3E}}%
\begin{equation}
\mathcal{L}_{ST}^{\left( H\right) }=\int d^{4}\theta _{{}}H\left( 
\boldsymbol{\Phi ,\bar{\Phi}};\boldsymbol{L}\right)
\end{equation}

\subsubsection{Scalar potential}

The scalar potential of the ADJ model (\ref{TS}-\ref{ST}) has two
contributions as follows\textrm{\ }%
\begin{equation}
\mathcal{V}_{sca}=\mathcal{V}_{gauge}+\mathcal{V}_{tens}
\end{equation}%
\textrm{\ }a contribution\textrm{\ }$\mathcal{V}_{gauge}$ coming from the
auxiliary fields\textrm{\ }$F^{X^{a}}$ and $D^{a}$\textrm{\ }of the gauge
multiplets; and another contribution $\mathcal{V}_{tens}$\textrm{\ }coming
from the auxiliary field\textrm{\ }$\boldsymbol{\Phi }$; seen that $%
\boldsymbol{L}$ has no auxiliary field. The $\mathcal{V}_{gauge}$
contribution reads explicitly as 
\begin{equation}
\mathcal{V}_{gauge}=H_{ab}\left( F^{X^{a}}\bar{F}^{\bar{X}^{b}}+\frac{1}{2}%
D^{a}D^{b}\right)
\end{equation}%
where $H_{ab}=\func{Im}\mathcal{F}_{ab}$ is the metric of the special-K\"{a}%
hler manifold with inverse $H^{ab}$.\ For the $\mathcal{V}_{tens}$
contribution, we have 
\begin{equation}
\mathcal{V}_{tens}=F^{\phi }G_{\phi \bar{\phi}}\bar{F}^{\bar{\phi}}
\end{equation}%
\textrm{\ }where $G_{\phi \bar{\phi}}$ is the analogue of metric $H_{ab}$
for the matter sector\textrm{.}\newline
Substituting the various auxiliary fields by their field equations, we
obtain the explicit expression of the full scalar potential of the model.
For the contribution $\mathcal{V}_{gauge}$, we have%
\begin{equation}
\mathcal{V}_{gauge}=H^{ab}\left( \frac{1}{2}r_{a}r_{b}+w_{a}\bar{w}%
_{b}\right)
\end{equation}%
with real $r_{a}$ and complex $w_{a}$ as follows 
\begin{equation}
\begin{tabular}{lll}
$r_{a}$ & $=$ & $g_{a}C+\frac{\xi _{a}}{2}$ \\ 
$w_{a}$ & $=$ & $g_{a}\phi +\frac{1}{4}e_{a}+\frac{i}{4\kappa _{c}}\mathcal{F%
}_{ac}$%
\end{tabular}%
\end{equation}%
Besides FI coupling constants, they depend on the degrees of freedom of the
tensor multiplet namely $C$ and $\phi $. The other contribution is given by 
\begin{equation}
\mathcal{V}_{tensor}=2\varrho ^{2}\left \vert gX\right \vert ^{2}
\end{equation}%
where we have set 
\begin{equation}
\varrho ^{2}=\sqrt{C^{2}+2\left \vert \phi \right \vert ^{2}}\qquad ,\qquad
gX=g_{1}X_{1}+g_{2}X_{2}
\end{equation}%
So the total scalar scalar potential reads as%
\begin{equation}
\mathcal{V}_{sca}=H^{ab}\left( \frac{1}{2}r_{a}r_{b}+w_{a}\bar{w}_{b}\right)
+2\left \vert gX\right \vert ^{2}\varrho ^{2}  \label{d}
\end{equation}%
Observe the two following features: first for $\varrho ^{2}=0$ and $w_{a}=0$%
, the scalar potential $\mathcal{V}_{sca}$ has a non zero value due to the
non-vanishing $\xi $ and hence ${\mathcal{N}}=2$ supersymmetry breaks down.
For $\varrho ^{2}=0$ and $\xi _{a}=e_{a}=0$, the scalar potential $\mathcal{V%
}_{sca}$ has as well a non zero value proportional to the magnetic FI
coupling as shown on the following expression%
\begin{equation}
\mathcal{V}_{sca}=\sum \frac{1}{16\kappa _{c}\kappa _{d}}\mathcal{F}%
_{ca}H^{ab}\mathcal{\bar{F}}_{bd}
\end{equation}%
The stationarity condition of the scalar potential with respect to the
various fields namely 
\begin{equation}
\frac{\partial \mathcal{V}_{sca}}{\partial X}=0\qquad ,\qquad \frac{\partial 
\mathcal{V}_{sca}}{\partial \phi }=0\qquad ,\qquad \frac{\partial \mathcal{V}%
_{sca}}{\partial C}=0
\end{equation}%
leads, for the case $\left \langle C^{2}+2\left \vert \phi \right \vert
^{2}\right \rangle =0$, to the following equation%
\begin{equation}
\mathcal{F}_{abc}\left[ F^{x^{b}}\left( \bar{F}^{\bar{x}^{c}}+\frac{1}{%
2\kappa _{c}}\right) +\frac{1}{2}D^{b}D^{c}\right] =0
\end{equation}%
leading to broken supersymmetric phase for the case where $\mathcal{F}%
_{abc}\neq 0$.

\subsection{Using ($\boldsymbol{Q}_{1}$, $\boldsymbol{Q}_{2}$) hypermultiplet%
}

Using duality transformations (\ref{dt}), one can also express the $\mathcal{%
N}=2$ tensor multiplet $\mathcal{T}^{\left( \mathcal{N}=2\right) }$ as a
hypermultiplet described by two $\mathcal{N}=1$ chiral superfields $%
\boldsymbol{Q}_{1}$ and $\boldsymbol{Q}_{2}$. In this realisation of $%
\mathcal{T}^{\left( \mathcal{N}=2\right) }$, the previous superspace density
(\ref{ST}) gets mapped to the equivalent expression 
\begin{eqnarray}
\mathcal{L}_{hyp} &=&\int d^{4}\theta _{{}}\left( \boldsymbol{\bar{Q}}%
_{1}e_{{}}^{g_{1}\boldsymbol{V}_{1}+g_{_{2}}\boldsymbol{V}_{2}}\boldsymbol{Q}%
_{1}+\boldsymbol{\bar{Q}}_{2}e_{{}}^{-g_{1}\boldsymbol{V}_{1}-g_{_{2}}%
\boldsymbol{V}_{2}}\boldsymbol{Q}_{2}\right)  \notag \\
&&+\int d^{2}\theta _{{}}\left( m+i\sqrt{2}g_{1}\boldsymbol{X}_{1}+i\sqrt{2}%
g_{2}\boldsymbol{X}_{2}\right) \boldsymbol{Q_{1}Q}_{2}  \label{hp} \\
&&-i\int d^{2}\theta _{{}}\left( \frac{g_{_{1}}}{\kappa _{1}}+\frac{g_{_{2}}%
}{\kappa _{2}}\right) \boldsymbol{Y}+hc  \notag
\end{eqnarray}%
where the superfields $\boldsymbol{Q}_{1}$ and $\boldsymbol{Q}_{2}$ carry
opposite charges under the $U_{1}\left( 1\right) \times U_{2}\left( 1\right) 
$ gauge symmetry; but $\boldsymbol{Y}$ playing the same role. \newline
The scalar potential of this superfield realisation of the the gauge theory
is given by%
\begin{equation}
\mathcal{V}_{sca}^{\prime }=H_{ab}\left[ F^{x^{a}}\bar{F}^{\bar{x}^{b}}+%
\frac{1}{2}D^{a}D^{b}\right] +G_{u\bar{v}}F^{q^{u}}\bar{F}^{\bar{q}^{\bar{v}%
}}
\end{equation}%
It has the same form as (\ref{d})%
\begin{equation}
\mathcal{V}_{sca}^{\prime }=\frac{1}{2}r_{a}^{\prime }H^{ab}r_{b}^{\prime
}+w_{a}^{\prime }H^{ab}\text{ }\bar{w}_{b}^{\prime }+\varrho ^{\prime
2}\left \vert m+\sqrt{2}ig_{a}x^{a}\right \vert ^{2}  \label{sp}
\end{equation}%
but now with the dual expressions%
\begin{equation}
\begin{tabular}{lll}
$r_{a}^{\prime }$ & $=$ & $-g_{a}\left( \left \vert q^{1}\right \vert
^{2}-\left \vert q^{2}\right \vert ^{2}\right) +\frac{\xi _{a}}{2}$ \\ 
$w_{a}^{\prime }$ & $=$ & $\frac{\sqrt{2}g_{a}}{i}q^{1}q^{2}+\frac{1}{4}%
e_{a}+\frac{i}{4\kappa _{c}}\mathcal{F}_{ac}$%
\end{tabular}%
\end{equation}%
and%
\begin{equation}
\varrho ^{\prime 2}=\left \vert q^{1}\right \vert ^{2}+\left \vert
q^{2}\right \vert ^{2}
\end{equation}%
The properties of the scalar potential (\ref{sp}), including the description
of the two scale breakings of $\mathcal{N}=2$ supersymmetry, have been
explicitly studied in \textrm{\cite{2D}}.

\section{ADJ model and tadpole anomaly}

In this section, we study a D3 brane realisation of ADJ theory and the
interpretation of partial supersymmetric breaking in terms of\ 3-forms
fluxes through 3-cycles in CY3. This brane realisation has been succinctly
presented in the introduction section; here we use results on type IIB
string on local CY3s to describe the underlying geometry and the nature of H$%
_{3}^{RR}$, H$_{3}^{NS}$ fluxes behind $\mathcal{N}=2$ ADJ model. To reach
this goal, we first examine the geometric property the linear combinations
of abelian gauge superfields; then we study the geometric derivation of the $%
\mathcal{N}=2$ ADJ condition and its $\mathcal{N}=1$ deformation given by (%
\ref{def}); and after we give the explicit relationship between ADJ
condition and D3 tadpole cancellation anomaly in type IIB.

\subsection{$\mathcal{N}=2$ ADJ model and 3-cycles in CY3}

In the effective $\mathcal{N}=2$ supersymmetric $U\left( 1\right) \times
U\left( 1\right) $ gauge model, the superspace lagrangian density $\mathcal{L%
}_{{\small U(1)}^{{\small 2}}}^{\mathcal{N}=2}$ depends, in addition to the
single tensor multiplet $\mathcal{T}^{\mathcal{N}=2}=\left( \boldsymbol{L},%
\boldsymbol{\Phi },\boldsymbol{Y}\right) $, on two $\mathcal{N}=2$ abelian
gauge multiplets $\mathcal{W}_{1,2}^{\mathcal{N}=2}=$ $\left( \boldsymbol{X}%
_{1},\boldsymbol{V}_{1}\right) $, $\left( \boldsymbol{X}_{2},\boldsymbol{V}%
_{2}\right) $.

\subsubsection{Linear combinations}

By an inspection of the superspace density (\ref{TS}), one notices that $%
\mathcal{L}_{{\small U(1)}^{{\small 2}}}^{\mathcal{N}=2}$ depends on the
following superfield linear combinations%
\begin{equation}
\begin{tabular}{lll}
$\boldsymbol{V}$ & $=$ & $g_{1}\boldsymbol{V}_{1}+g_{_{2}}\boldsymbol{V}_{2}$
\\ 
$\boldsymbol{V}^{\prime }$ & $=$ & $\xi _{1}\boldsymbol{V}_{1}+\xi _{_{2}}%
\boldsymbol{V}_{2}$%
\end{tabular}
\label{LC}
\end{equation}%
and%
\begin{eqnarray}
\boldsymbol{X} &=&g_{1}\boldsymbol{X}_{1}+g_{_{2}}\boldsymbol{X}_{2}  \notag
\\
\boldsymbol{X}^{\prime } &=&e_{1}\boldsymbol{X}_{1}+e_{_{2}}\boldsymbol{X}%
_{2} \\
\frac{\partial \mathcal{F}}{\partial X} &=&\frac{1}{\boldsymbol{\kappa }_{1}}%
\frac{\partial \mathcal{F}}{\partial X_{1}}+\frac{1}{\boldsymbol{\kappa }_{2}%
}\frac{\partial \mathcal{F}}{\partial X_{2}}  \notag
\end{eqnarray}%
\begin{equation*}
\end{equation*}%
These superfield combinations may a priori be extended to any number n of $%
\mathcal{N}=2$ gauge multiplets $\mathcal{W}_{a}^{\mathcal{N}=2}=\left( 
\boldsymbol{X}_{a},\boldsymbol{V}_{a}\right) $ as follows%
\begin{eqnarray}
\boldsymbol{V} &=&\dsum \limits_{a=1}^{n}g_{a}\boldsymbol{V}_{a}  \label{BL2}
\\
\boldsymbol{X} &=&\dsum \limits_{a=1}^{n}g_{a}\boldsymbol{X}_{a}  \label{BL1}
\end{eqnarray}%
where the $g_{a}$'s are gauge coupling constants associated with each
abelian $U_{a}\left( 1\right) $ gauge multiplet $\left( \boldsymbol{X}_{a},%
\boldsymbol{V}_{a}\right) $. Similar relations can be written down for $%
\boldsymbol{V}^{\prime },$ $\boldsymbol{X}^{\prime }$ and $\frac{\partial 
\mathcal{F}}{\partial X}$. However, because of ADJ constraint relation; the
generalisation of the condition (\ref{rc}) to arbitrary U$\left( 1\right)
^{n}$ gauge symmetry is valid provided $n\geq 2$ as in (\ref{cst}). The
restriction to the particular $n=1$ case leads to singular relation%
\begin{equation*}
\frac{g_{1}}{\boldsymbol{\kappa }_{1}}=0
\end{equation*}
requiring $g=0$ for $\boldsymbol{\kappa }\neq 0$. To overcome this
difficulty; one may resolve the $\frac{g_{1}}{\boldsymbol{\kappa }_{1}}=0$
singularity by deforming it like $\nu +\frac{g}{\boldsymbol{\kappa }}=0$;
this leads to $g=-\boldsymbol{\kappa }\nu $; however remembering the
property of eqs(\ref{ob}-\ref{def}), one learns that the deformation by $\nu 
$ breaks explicitly $\mathcal{N}=2$ supersymmetry down to $\mathcal{N}=1$.%
\newline
To see the meaning of the linear combinations (\ref{LC}--\ref{BL1}) as well
as the interpretation of the deformation 
\begin{equation}
\nu +\dsum \limits_{a=1}^{n}\frac{g_{a}}{\boldsymbol{\kappa }_{a}}=0
\label{nu}
\end{equation}%
we need to go beyond 4d space time by thinking of:

\begin{itemize}
\item the $\mathcal{N}=2$ supersymmetric $U\left( 1\right) ^{n}$ gauge model
as a part of an effective theory following from type IIB string compactified
on a local CY3; and

\item the constraint relation (\ref{nu}) as corresponding to the D3 tapole
anomaly \cite{1G}%
\begin{equation}
\frac{1}{2\kappa _{10}^{2}T_{3}}\int_{CY3}\hat{H}_{3}^{RR}\wedge \hat{H}%
_{3}^{NS}+N_{D3}=0
\end{equation}%
where the 3-form gauge field strengths $\hat{H}_{3}^{RR}$, $\hat{H}_{3}^{NS}$%
; and the numbers $T_{3}$ and $N_{D3}$ will be introduced later on.
\end{itemize}

\  \  \  \  \  \newline
To be explicit, we study in what follows the derivation of the linear
combinations $\sum_{a}\xi _{a}\boldsymbol{V}^{a}$ and $\sum_{a}g_{a}%
\boldsymbol{V}^{a}$ from type IIB string compactification on a Calabi -Yau
threefold $\mathcal{Z}_{3}$ with Kahler 2-form $J_{2}$ and complex
holomorphic 3-form $\Omega _{3}$. Then, we turn to the derivation of the
linear combinations $\sum_{a}e_{a}\boldsymbol{X}^{a}$ and $\sum_{a}\frac{1}{%
\boldsymbol{\kappa }_{a}}\frac{\partial \mathcal{F}}{\partial X^{a}}$
concerning the chiral superfields.

\subsubsection{Kahler sector}

To derive the two linear combinations involving the gauge multiplet namely
the $\boldsymbol{V}=\sum_{a}\xi _{a}\boldsymbol{V}^{a}$, depending on FI
coupling constants $\xi _{a}$, and the $\boldsymbol{V}^{\prime
}=\sum_{a}g_{a}\boldsymbol{V}^{a}$ involving gauge coupling constants $g_{a}$%
, it is interesting\ to start by describing the $\theta $- expansions of
these combinations. Focussing on the 4-vector $\upsilon _{\mu }^{a}$ field
components of the linear sum of gauge superfields $\boldsymbol{V}^{a}$ which
expands in $\theta $- series as 
\begin{eqnarray}
\boldsymbol{V} &=&\theta _{{}}\sigma ^{\mu }\bar{\theta}_{{}}\left( \upsilon
_{\mu }\right) +\theta _{{}}^{2}\bar{\theta}_{{}}^{2}\left(
\sum_{a=1}^{n}\chi _{a}D^{a}\right)  \notag \\
&&+\frac{i}{\sqrt{2}}\theta _{{}}^{2}\bar{\theta}_{{}}\left(
\sum_{a=1}^{n}\chi _{a}\bar{\lambda}^{a}\right) -\frac{i}{\sqrt{2}}\bar{%
\theta}_{{}}^{2}\theta _{{}}\left( \sum_{a=1}^{n}\chi _{a}\lambda ^{a}\right)
\end{eqnarray}%
with 
\begin{equation}
\upsilon _{\mu }=\sum_{a=1}^{n}\chi _{a}\upsilon _{\mu }^{a}\qquad ,\qquad
\chi _{a}=\xi _{a},\text{ }g_{a}
\end{equation}%
Obviously for the case $\chi _{a}=\xi _{a}$, the contribution to ADJ model
is given by the D- term $\theta _{{}}^{2}\bar{\theta}_{{}}^{2}\left(
\sum_{a=1}^{n}\xi _{a}D^{a}\right) $; and the interpretation of $\xi _{a}$'s
may be obtained by computing the field eqs of the auxiliary $D^{a}$ fields.
However, we can reach the same result by looking for the derivation of this
quantity from superstring compactification.

\  \  \  \ 

\emph{i) FI coupling constants}\newline
As a first step toward the $\xi _{a}\boldsymbol{V}^{a}$'s, we use the 4d
space time language of 1-form gauge field potentials $V_{1}^{a}=\upsilon
_{\mu }^{a}dx^{\mu }$ to rewrite the gauge component field linear
combination $\sum \xi _{a}\upsilon _{\mu }^{a}$ as follows%
\begin{equation}
\left( \sum_{a=1}^{n}\xi _{a}\upsilon _{\mu }^{a}\right) dx^{\mu
}=\sum_{a=1}^{n}\xi _{a}V_{1}^{a}\equiv V_{1}  \label{pt}
\end{equation}%
So $\xi _{a}\upsilon _{\mu }^{a}$ can be also viewed in terms of a linear
combination of the 1-form gauge field potentials $V_{1}^{a}$. The next step
is to transform above (\ref{pt}) into an integral over full dimensions of
the Calabi-Yau threefolds; this is achieved by thinking about the 1-form
gauge field $V_{1}^{a}$ in 4d space time as due to a 4- form gauge potential 
$\hat{C}_{4}$ of a D3 brane living in 10d space time%
\begin{equation}
\hat{C}_{4}=\frac{1}{4!}C_{MNPQ}\text{ }d_{{}}\hat{x}^{M}\wedge d_{{}}\hat{x}%
^{N}\wedge d_{{}}\hat{x}^{P}\wedge d_{{}}\hat{x}^{Q}
\end{equation}%
but with three directions wrapping the compact 3-cycles $\left[ A_{a}\right] 
$ of the local CY3 as follows 
\begin{equation}
V_{1}^{a}=\frac{1}{2\pi \alpha ^{\prime }}\dint \nolimits_{CY3}\hat{C}%
_{4}\wedge \mathrm{\beta }^{a}  \label{cb}
\end{equation}%
with harmonic 3-form $\mathrm{\beta }^{a}$ belonging to $H^{\left(
2,1\right) }\left( CY3,R\right) \oplus H^{\left( 1,2\right) }\left(
CY3,R\right) $. Putting this relation back into (\ref{pt}), we end with%
\begin{equation}
\sum_{a=1}^{n}\xi _{a}V_{1}^{a}=\frac{1}{2\pi \alpha ^{\prime }}\dint
\nolimits_{CY3}\hat{C}_{4}\wedge dJ_{2}  \label{e}
\end{equation}%
with $dJ_{2}$ standing for 3-form obtained by complex deformation of Kahler
2-form $J_{2}$ of the CY3 \textrm{\cite{CY3,5EA,5EB,5EC}} 
\begin{equation}
dJ_{2}=\sum_{a=1}^{n}\xi _{a}\mathrm{\beta }^{a}\qquad ,\qquad \xi _{a}=%
\frac{1}{2}\left( \dint \nolimits_{\left[ B^{a}\right] }dJ_{2}+hc\right)
\label{j}
\end{equation}%
where 3-cycle $\left[ B^{a}\right] $ is the dual of $\left[ A_{a}\right] $
in the CY3.\ Recall that the pair $\left[ A_{a}\right] $ and $\left[ B^{a}%
\right] $ form a symplectic basis of 3-cycles in the homology group of the
CY3; they are in 1:1 correspondence with the the 3-form harmonic basis $%
\left( \mathrm{\alpha }_{a}^{{}},\mathrm{\beta }^{a}\right) $ of the
cohomology group.

\  \  \  \  \  \ 

\emph{ii) gauge coupling constants}\newline
To derive the linear combination $\sum_{a=1}^{n}g_{a}\upsilon _{\mu }^{a}$
and the expression of the gauge coupling constants $g_{a}$, we use the 4d
Chern-Simons interaction $\mathcal{L}_{CS}^{4d}$ between the gauge
potentials $\upsilon _{\mu }^{a}$ and the the antisymmetric field strength $%
H_{\nu \rho \sigma }$, 
\begin{equation}
\mathcal{L}_{CS}^{4d}=\frac{1}{3!}\left( \sum_{a=1}^{n}g_{a}\upsilon _{\mu
}^{a}H_{\nu \rho \sigma }\right) \varepsilon ^{\mu \nu \rho \sigma }
\end{equation}%
that we rewrite, by using wedge product $V_{1}^{a}\wedge H_{3}$ of 4d space
time 1- and 3- forms like 
\begin{equation}
\mathcal{S}_{CS}^{4d}=\dint \nolimits_{M^{4}}\left(
\sum_{a=1}^{n}g_{a}V_{1}^{a}\right) \wedge H_{3}
\end{equation}%
But seen that in type IIB string, we have two kinds of 3-forms $H_{3}^{RR}$
and $\hat{H}_{3}^{NS}$; then we can think of the 4d Chern-Simons action $%
\mathcal{S}_{CS}^{4d}$ as resulting from the following 10d expression%
\begin{equation}
\mathcal{S}_{CS}^{10d}=\dint \nolimits_{M^{10}}\hat{C}_{4}\wedge \hat{H}%
_{3}^{NS}\wedge \hat{H}_{3}^{RR}
\end{equation}%
This relation leads in general to two kinds of 4d space time contributions;
one involving 4d space time 3-form $H_{3}^{RR}$ and the other the 4d 3-form $%
H_{3}^{NS}$ as follows 
\begin{equation}
\dint \nolimits_{M^{4}}\left( \dint \nolimits_{CY3}\hat{C}_{4}\wedge \hat{H}%
_{3}^{NS}\right) \wedge H_{3}^{RR}-\dint \nolimits_{M^{4}}\left( \dint
\nolimits_{CY3}\hat{C}_{4}\wedge \hat{H}_{3}^{RR}\right) \wedge H_{3}^{NS}
\label{2L}
\end{equation}%
If we restrict to first contribution and comparing with (\ref{cb}), we end
with the following expression for the gauge coupling constants%
\begin{equation}
g_{a}=\frac{1}{2\pi \alpha ^{\prime }}\dint \nolimits_{CY3}\mathrm{\alpha }%
_{a}\wedge H_{3}^{NS}
\end{equation}%
that reads also as follows 
\begin{equation}
g_{a}=\frac{1}{2\pi \alpha ^{\prime }}\dint \nolimits_{B^{a}}^{\Lambda
_{0}}H_{3}^{NS}  \label{ns}
\end{equation}%
with $\alpha ^{\prime }$ the string constant and where $\Lambda _{0}$ is a
cut off playing the role of running scale of the well known renormalisation
group equation.

\  \  \  \  \  \newline
To conclude, the linear combinations of the $\mathcal{N}=2$ gauge multiplets
used in ADJ model have the following geometric interpretation in type IIB
string on local CY3 
\begin{eqnarray}
\sum_{a=1}^{n}\xi _{a}V_{1}^{a} &=&\frac{1}{2\pi \alpha ^{\prime }}\dint
\nolimits_{CY3}\hat{C}_{4}\wedge dJ_{2} \\
\dsum \limits_{a\geq 1}g_{a}V_{1}^{a} &=&\frac{1}{4\pi ^{2}\alpha ^{\prime 2}%
}\dint \nolimits_{CY3}\hat{C}_{4}\wedge H_{3}^{NS}
\end{eqnarray}

\subsubsection{Chiral sector}

The $\boldsymbol{X}^{a}$'s are chiral superfields with $\theta $- components
given by complex scalar fields $\left. \boldsymbol{X}^{a}\right \vert
_{\theta =0}=X^{a}$; Weyl spinors $\left. D_{\alpha }\boldsymbol{X}%
^{a}\right \vert _{\theta =0}=\psi _{\alpha }^{a}$ and auxiliary fields $%
\left. \frac{1}{2}D^{\alpha }D_{\alpha }\boldsymbol{X}^{a}\right \vert
_{\theta =0}=F^{a}$. The linear combination $\sum e_{a}\boldsymbol{X}^{a}$
is a chiral superfield \ 
\begin{equation}
\sum_{a\geq 1}e_{a}\boldsymbol{X}^{a}=X+\theta .\left( \sum_{a\geq
1}e_{a}\psi ^{a}\right) +\theta ^{2}\left( \sum_{a\geq 1}e_{a}F^{a}\right)
\label{px}
\end{equation}%
with leading $\theta $- component given by a similar relation to the
superfield one namely 
\begin{equation}
X=\sum_{a\geq 1}e_{a}X^{a}  \label{cx}
\end{equation}%
A similar relation is valid for the magnetic FI combination 
\begin{equation}
\frac{1}{\kappa }\frac{\partial \mathcal{F}}{\partial X}=\sum_{a\geq 1}\frac{%
1}{\kappa _{a}}\frac{\partial \mathcal{F}}{\partial X^{a}}
\end{equation}%
In the embedding of ADJ model into 10d type IIB string compactification on
CY3, the complex scalars $X^{a}$ and $\mathcal{F}_{a}$ describe the
expansion modes of the holomorphic 3- form $\Omega _{3}$ over the 3-form
harmonic basis $\left( \mathrm{\alpha }_{a},\mathrm{\beta }^{a}\right) $ of
the local CY3%
\begin{equation}
\Omega _{3}=X^{a}\mathrm{\alpha }_{a}-\mathcal{F}_{a}\mathrm{\beta }^{a}
\end{equation}%
with%
\begin{equation}
X^{a}=\dint \nolimits_{CY3}\Omega _{3}\wedge \mathrm{\beta }^{a}\qquad
,\qquad \mathcal{F}_{a}=\dint \nolimits_{CY3}\Omega _{3}\wedge \mathrm{%
\alpha }_{a}  \label{xa}
\end{equation}%
The electric FI $e_{a}$ and magnetic $\frac{1}{\kappa _{a}}$ coupling
constants are obtained by equating the superpotential%
\begin{equation}
W\left( X\right) =e_{a}X^{a}-\frac{1}{\kappa _{a}}\mathcal{F}_{a}
\end{equation}%
with the expression of $W\left( X\right) $, build out of $%
G_{3}=H_{3}^{RR}-\tau H_{3}^{NS}$ and the complex holomorphic 3-form namely 
\textrm{\cite{GVW,1G}}, 
\begin{equation}
W\left( X\right) \sim \dint \nolimits_{CY3}\Omega _{3}\wedge G_{3}
\end{equation}%
We obtain%
\begin{equation}
G_{3}=e_{a}\mathrm{\beta }^{a}-\frac{1}{\kappa _{a}}\mathrm{\alpha }_{a}
\end{equation}%
So, we have%
\begin{equation}
e_{a}=\dint \nolimits_{CY3}\mathrm{\alpha }_{a}\wedge G_{3}\qquad ,\qquad 
\frac{1}{\kappa _{a}}=\dint \nolimits_{CY3}G_{3}\wedge \mathrm{\beta }^{a}
\label{kap}
\end{equation}

\subsection{Deriving ADJ constraint from type IIB string}

In ADJ model, the condition (\ref{cst}) is intimately related with the real
4-form $C_{4}$ with antisymmetric gauge potential field $C_{\mu \nu \rho
\sigma }$ as in (\ref{Y}) and directions filling the 4d space time
dimensions 
\begin{eqnarray}
C_{4} &=&C_{\mu \nu \rho \sigma }dx^{\mu }\wedge dx^{\nu }\wedge dx^{\rho
}\wedge dx^{\sigma }  \notag \\
F_{5} &=&0
\end{eqnarray}%
This 4-form gauge field potential cannot have field strength $F_{5}=dC_{4}$
in 4d space time; and thus it should be treated as an auxiliary field as
done by ADJ theory. Notice that this is a constant field that may be ignored
by setting it to zero; but this corresponds to a particular solution since
in general it reads like 
\begin{eqnarray}
C_{\mu \nu \rho \sigma } &=&\Delta \varepsilon _{\mu \nu \rho \sigma }\qquad
,\qquad \Delta =cst  \notag \\
d\Delta &=&0  \label{del}
\end{eqnarray}%
and captures a constraint relation that we study below. Observe that $\Delta 
$ scales as mass$^{2}$; and so the scaling mass dimension of $C_{4}$ is mass$%
^{-2}$.

\subsubsection{Tadpole anomaly}

To get more insight into the meaning of the $\mathcal{N}=2$ ADJ condition $%
\sum_{a}\frac{g_{a}}{\boldsymbol{\kappa }_{a}}=0$ and its $\mathcal{N}=1$
deformation $\nu +\sum_{a}\frac{g_{a}}{\boldsymbol{\kappa }_{a}}=0$, one has
to go beyond 4d space time where the gauge potential $C_{4}$ is no longer
constant%
\begin{equation}
\left. F_{5}\right \vert _{4d}=\left. dC_{4}\right \vert _{4d}=0\qquad
,\qquad \left. \hat{F}_{5}\right \vert _{10d}=\left. d\hat{C}_{4}\right
\vert _{10d}\neq 0
\end{equation}%
In a higher dimension D space time compactified on real $\left( D-4\right) $
manifold, $M^{D}=M^{1,3}\times M^{\left( D-4\right) }$, some of the
directions of the extended 4-form field $\hat{C}_{4}$ can wrap dimensions in
the internal space allowing as consequence Chern-Simons couplings between $%
\hat{C}_{4}$ and other p$_{i}$-forms $\hat{F}_{p_{i}}$ living in $M^{D}$;
for example $\mathcal{S}_{_{CS}}^{D}\sim \dint_{M^{D}}\hat{C}_{4}\wedge \hat{%
F}_{D-4}$. This is exactly what happens in the case of type IIB strings
compactified on Calabi- Yau threefolds where the 4- form gauge potential $%
\hat{C}_{4}$, sourced by D3 brane, appears naturally and where couplings
with other p-form gauge field potentials are possible. In 10d type IIB
theory, the Chern Simons coupling reads as follows 
\begin{equation}
\mathcal{S}_{_{CS}}^{10d}\sim \dint_{M^{1,9}}\hat{C}_{4}\wedge \hat{H}%
_{3}^{RR}\wedge \hat{H}_{3}^{NS}  \label{SC}
\end{equation}%
where the 3- forms $\hat{H}_{3}^{NS}$ and $\hat{H}_{3}^{RR}$ are the gauge
field strengths of the antisymmetric $\hat{B}_{\mu \nu }^{NS}$ and $\hat{B}%
_{\mu \nu }^{RR}$ gauge potentials of type IIB strings. The field equation
of the 5-form field strength $\hat{F}_{5}$ of the $\hat{C}_{4}$ gauge
potential is given by 
\begin{equation}
dF_{5}=\hat{H}_{3}^{NS}\wedge \hat{H}_{3}^{RR}+2\kappa _{10}^{2}T_{3}\varrho
_{3}^{loc}  \label{5F}
\end{equation}%
with $T_{3}=\frac{1}{\left( 2\pi \right) ^{3}\alpha ^{\prime 2}}$ the D3-
brane tension, $2\kappa _{10}^{2}=\left( 2\pi \right) ^{7}\alpha ^{\prime 4}$%
; and where $\varrho _{3}^{loc}$- form stands for the D3 charge density due
to localized sources including D7-branes or O3 planes and also of mobile
D3-branes \textrm{\cite{1G}}; see also\textrm{\  \cite{2G,3G}}%
\begin{equation}
\dint \nolimits_{CY3}\varrho _{3}^{loc}=N_{D3}  \label{den}
\end{equation}%
From 4d space time view, the integration of the Chern-Simons coupling (\ref%
{SC}) on the internal coordinates 
\begin{equation}
\mathcal{S}_{_{CS}}^{4d}=\dint_{M^{1,3}}\left( \dint \nolimits_{CY3}\hat{C}%
_{4}\wedge \hat{H}_{3}^{RR}\wedge \hat{H}_{3}^{NS}\right)
\end{equation}%
leads to various terms that can be organised into three block terms 
\begin{equation}
\mathcal{S}_{_{CS}}^{4d}=\dint_{M^{1,3}}\left( \mathcal{L}_{0}^{4d}+\mathcal{%
L}_{1}^{4d}+\mathcal{L}_{2}^{4d}\right)  \label{CS}
\end{equation}%
with 
\begin{eqnarray}
\mathcal{L}_{0}^{4d} &\sim &\left( \dint \nolimits_{CY3}\hat{H}%
_{3}^{RR}\wedge \hat{H}_{3}^{NS}\right) \wedge C_{4}  \label{tf} \\
\mathcal{L}_{1}^{4d} &\sim &\left( \dint \nolimits_{CY3}\hat{C}_{4}\wedge 
\hat{H}_{3}^{RR}\right) \wedge H_{3}^{NS}  \label{te} \\
\mathcal{L}_{2}^{4d} &\sim &-\left( \dint \nolimits_{CY3}\hat{C}_{4}\wedge 
\hat{H}_{3}^{NS}\right) \wedge H_{3}^{RR}  \label{th}
\end{eqnarray}%
where un-hatted fields refer to 4d space time fields. Notice that the two
last terms are precisely the ones given by eqs(\ref{2L}), they contribute to
Chern-Simons couplings in 4d space time. \newline
The remaining term (\ref{tf}), involving the integral of $\hat{H}%
_{3}^{RR}\wedge \hat{H}_{3}^{NS}$ through full CY3; is a topological term
describing the flux $\Phi _{flux}$ of the 6-form $\hat{H}_{3}^{RR}\wedge 
\hat{H}_{3}^{NS}$ through the CY3 
\begin{equation}
\Phi _{flux}\sim \dint \nolimits_{CY3}\hat{H}_{3}^{RR}\wedge \hat{H}_{3}^{NS}
\label{x}
\end{equation}%
For compact CY3, this flux has no where to go and so has to vanish; but as
we will see in a moment it may be compensated by D3- brane charges coming
from local sources. So the contribution of this term to 4d space time
lagrangian density reads as 
\begin{equation}
\frac{\Phi _{flux}}{24}\  \varepsilon ^{\mu \nu \rho \sigma }C_{\mu \nu \rho
\sigma }=\frac{\Phi _{flux}}{24}\Delta
\end{equation}%
where we have used (\ref{del}). This term appears linearly in the ADJ
lagrangian density; so it captures a constraint equation requiring the flux $%
\Phi _{flux}$ to vanish as noticed above. This constraint can be then
interpreted as nothing but the vanishing condition given by the integral of (%
\ref{5F}) on CY3 for the particular case $\varrho _{3}^{loc}=0$. For a non
zero $\varrho _{3}^{loc}$ density of D3- brane charges as in eq(\ref{den}),
the tadpole vanishing condition reads as 
\begin{equation}
\frac{1}{2\kappa _{10}^{2}T_{3}}\int_{CY3}\hat{H}_{3}^{RR}\wedge \hat{H}%
_{3}^{NS}+N_{D3}=0  \label{TD}
\end{equation}%
and the contribution to the 4d space time lagrangian density gets modified
like 
\begin{equation}
\mathcal{L}_{0}^{4d}=\frac{\Delta }{24}\text{ }\left( \Phi
_{flux}+N_{D3}\right)  \label{45}
\end{equation}%
The presence of the $N_{D3}$ flux breaks explicitly $\mathcal{N}=2$
supersymmetry down to $\mathcal{N}=1$; $N_{D3}$ flux plays the same role as
the parameter $\nu $ in the deformed constraint eq(\ref{nu}).

\subsubsection{Revisiting $\mathcal{N}=2$ ADJ condition and its deformation}

Here, we give the general expression of the condition of tadpole
cancellation in terms of the $H_{3}^{RR}$ and $H_{3}^{NS}$ fluxes. If
assuming that the CY3 is compact, then we have%
\begin{eqnarray}
\frac{1}{2\pi \alpha ^{\prime }}\int_{CY3}H_{3}^{RR}\wedge \mathrm{\beta }%
^{a} &=&p_{a}\   \notag \\
\frac{1}{2\pi \alpha ^{\prime }}\int_{CY3}\mathrm{\alpha }_{a}\wedge
H_{3}^{RR} &=&p_{n+a}  \label{g}
\end{eqnarray}%
and%
\begin{eqnarray}
\frac{1}{2\pi \alpha ^{\prime }}\int_{CY3}H_{3}^{NS}\wedge \mathrm{\beta }%
^{a} &=&q_{a}\   \notag \\
\frac{1}{2\pi \alpha ^{\prime }}\int_{CY3}\mathrm{\alpha }_{a}\wedge
H_{3}^{NS} &=&q_{n+a}
\end{eqnarray}%
From these relations, we learn%
\begin{equation}
\frac{1}{4\pi ^{2}\alpha ^{\prime 2}}\int_{CY3}\hat{H}_{3}^{NS}\wedge \hat{H}%
_{3}^{RR}=q_{A}\Omega ^{AB}p_{B}  \label{ge}
\end{equation}%
where the quantised vectors are as%
\begin{equation}
p_{A}=\left( 
\begin{array}{c}
p_{a} \\ 
p_{a+n}%
\end{array}%
\right) \qquad ,\qquad q_{A}=\left( 
\begin{array}{c}
q_{b} \\ 
q_{b+n}%
\end{array}%
\right)
\end{equation}%
and where $\Omega ^{AB}$ is the usual $2n\times 2n$ symplectic matrix 
\begin{equation}
\Omega ^{AB}=\left( 
\begin{array}{cc}
0 & I_{n\times n} \\ 
-I_{n\times n} & 0%
\end{array}%
\right)
\end{equation}%
Putting back into (\ref{TD}), the tadpole condition becomes%
\begin{equation}
\nu +4\pi ^{2}\alpha ^{\prime 2}\text{ }q_{A}\Omega ^{AB}p_{B}=0
\end{equation}%
with%
\begin{equation}
\nu =2\kappa _{10}^{2}T_{3}N_{D3}=\left( 2\pi \right) ^{4}\alpha ^{\prime
2}N_{D3}
\end{equation}%
The $\mathcal{N}=2$ ADJ constraint (\ref{cst}) may be recovered by requiring 
$N_{D3}=0;$ and choosing the vectors $p_{A}$ and $q_{A}$ like%
\begin{equation}
\frac{1}{\boldsymbol{\kappa }_{a}}=2\pi \alpha ^{\prime 2}q_{a}\qquad
,\qquad g_{a}=2\pi p_{a+n}  \label{eg}
\end{equation}%
the other $p_{a}$ and $q_{a+n}$ are set to zero.

\section{Conclusion and comments}

In this paper, we have studied a D3 brane realisation of partial breaking of 
$\mathcal{N}=2$ supersymmetry in ADJ model of ref \textrm{\cite{2D}}. This
is a particular 4d $\mathcal{N}=2$ supersymmetric $U\left( 1\right) \times
U\left( 1\right) $ gauge model describing the coupling of the $\mathcal{N}=2$
gauge multiplets $\mathcal{W}_{1}^{\left( \mathcal{N}=2\right) },$ $\mathcal{%
W}_{2}^{\left( \mathcal{N}=2\right) }$ with a single tensor $\mathcal{T}%
^{\left( \mathcal{N}=2\right) }$; it is also the leading model in the family
of effective 4d $\mathcal{N}=2$ $U\left( 1\right) ^{n}$ gauge theory
describing the dynamics of the multiplet $\mathcal{T}^{\left( \mathcal{N}%
=2\right) }$ coupled to n $\mathcal{N}=2$ Maxwell gauge superfield strengths 
$\mathcal{W}_{1}^{\left( \mathcal{N}=2\right) },...,$ $\mathcal{W}%
_{n}^{\left( \mathcal{N}=2\right) }$ with $n\geq 2$. The coupling between
gauge and matter superfields is of Chern-Simons type; it reads in $\mathcal{N%
}=2$ chiral superspace as follows 
\begin{equation}
\mathcal{L}_{CS}=\dint d^{8}\theta \mathcal{T}^{\left( \mathcal{N}=2\right)
}\left( \sum_{a=1}^{n}g_{a}\mathcal{W}_{a}^{\left( \mathcal{N}=2\right)
}\right)  \label{c1}
\end{equation}%
One of the basic constraint equations in this 4d effective $\mathcal{N}=2$
supersymmetric $U\left( 1\right) ^{n}$ gauge theory, formulated in $\mathcal{%
N}=1$ superspace with lagrangian density $\mathcal{L}=\mathcal{L}_{gauge}+%
\mathcal{L}_{ST}+\mathcal{L}_{CS},$ is the one given by $\frac{\delta 
\mathcal{L}}{\delta Y}=0$; the superfield equation of the auxiliary $%
\boldsymbol{Y}$ whose full contribution comes from the $\mathcal{L}_{CS}$
term namely 
\begin{equation}
\frac{\delta \mathcal{L}_{CS}}{\delta Y}=0
\end{equation}%
Because of linear dependence of $\mathcal{L}_{CS}$ into the superfield $%
\boldsymbol{Y}$, the above superfield equation turns into the constraint $%
\sum_{a=1}^{n}\frac{g_{a}}{\kappa _{a}}=0$. This condition relates the gauge
coupling constants $g_{a}$ to magnetic FI couplings $\frac{1}{\kappa _{a}}$;
and, upon computing energy of the ground state of the model, one also
obtains links between partial supersymmetry breaking scales and the $g_{a},$ 
$\kappa _{a}$ constant parameters as done in \textrm{\cite{2D}}. \newline
In our brane realisation, the ADJ model is represented by D3 branes wrapping
3-cycles in type IIB on local CY3 in presence of non trivial fluxes of the
3- forms gauge field strengths $H_{3}^{RR}$ and $H_{3}^{NS}$. In this
picture, the Chern-Simons coupling (\ref{c1}) is associated with a
particular term in the Calabi-Yau compactification of 10- $\dim $ field
action 
\begin{equation}
\mathcal{S}_{_{CS}}^{10d}\sim \dint_{M^{1,9}}\hat{C}_{4}\wedge \hat{H}%
_{3}^{RR}\wedge \hat{H}_{3}^{NS}
\end{equation}%
down to 4d space time; for details see eqs (\ref{SC}-\ref{th}). In this
brane representation of ADJ theory, the constraint relation $\sum_{a=1}^{n}%
\frac{g_{a}}{\kappa _{a}}=0$ is a particular realisation of the total flux
conservation condition $\Phi _{flux}=\frac{1}{2\kappa _{10}^{2}T_{3}}$ $%
\int_{CY3}\hat{H}_{3}^{NS}\wedge \hat{H}_{3}^{RR}=0$ which, by using the n-
dimensional symplectic homology basis of 3-cycles $\left( A^{a},B_{a}\right) 
$ of the CY3, reads in general like $\Phi _{flux}=q_{A}\Omega ^{AB}p_{B}$
with $p_{A}$ and $q_{B}$ as in eq(\ref{g} - \ref{ge}).%
\begin{equation}
\Phi _{flux}=\left( q_{a},q_{a+n}\right) \left( 
\begin{array}{cc}
0 & \delta _{ab} \\ 
-\delta _{ab} & 0%
\end{array}%
\right) \left( 
\begin{array}{c}
p_{b} \\ 
p_{b+n}%
\end{array}%
\right)
\end{equation}%
By expanding the above relation as $\Phi _{flux}=q_{a}p_{a+n}-q_{a+n}p_{a}$;
it follows that the condition $\sum_{a=1}^{n}\frac{g_{a}}{\kappa _{a}}=0$ is
indeed a particular solution of vanishing $\Phi _{flux}=0$ given the choice
of fluxes as in eq(\ref{eg}). This flux choice corresponds to expanding
gauge field strengths $H_{3}$ on the basis $\left( \alpha _{a},\beta
^{a}\right) $ of 3-forms, dual to 3-cycles basis, like $H_{3}^{NS}=q^{a}%
\alpha _{a}$ and $H_{3}^{RR}=p_{a+n}\beta ^{a}$. Geometrically, these values
correspond to the local Calabi-Yau picture where the 3-cycles $B_{a}$ are
taken in the non compact space approximation. In the case $n=1$; describing
the special $\mathcal{N}=2$ $U\left( 1\right) $ gauge model, the CY3 is the $%
T^{\ast }S^{3}$ conifold with compact $A$ given by the 3-sphere $S^{3}$ and
non compact $B\simeq \mathbb{R}^{3}$; and the above expression of the $\Phi
_{flux}$ flux reduces to 
\begin{equation}
\left( q_{1},0\right) \left( 
\begin{array}{cc}
0 & 1 \\ 
-1 & 0%
\end{array}%
\right) \left( 
\begin{array}{c}
0 \\ 
p_{2}%
\end{array}%
\right) =q_{1}p_{2}
\end{equation}%
The relationship between ADJ constraint and the flux $q_{A}\Omega ^{AB}p_{B}$
shows that $\sum_{a=1}^{n}\frac{g_{a}}{\kappa _{a}}=0$ is nothing but the
vanishing condition of the D3 tadpole anomaly in absence of D7 branes or O3
planes; the particular $n=1$ case is therefore anomalous. \newline
By taking into account D3 charge density due to localized sources D7-branes
or O3 planes; and by using a result from the study of \textrm{\cite{1G}},
the previous total flux conservation condition gets promoted to the
following relation 
\begin{equation}
\tilde{\Phi}_{flux}=\Phi _{flux}+N_{D3}=0
\end{equation}%
with $N_{D3}$ as in (\ref{den}). In this picture, the tadpole anomaly can be
lifted; but with the price of breaking explicitly $\mathcal{N}=2$
supersymmetry in the underlying effective field theory down to $\mathcal{N}%
=1 $. In this situation, the ADJ constraint $\sum_{a}\frac{g_{a}}{\kappa _{a}%
}=0 $ becomes deformed like $\nu +\sum_{a}\frac{g_{a}}{\kappa _{a}}=0$ as
shown by eq(\ref{45}).

\begin{acknowledgement}
This work is supported by URAC 09/CNRST; Saidi thanks ICTP- Trieste, Italie;
where a part of this work has been done.
\end{acknowledgement}

\section{Appendices}

Here we give two appendices; in appendix $\S $ $6.1$, we collect useful
tools on type IIB on CY3; and in appendix $\S $ $6.2$, we study properties
shared by supersymmetric and gauge transformations of the $\mathcal{N}=2$
gauge superfield $\mathcal{W}^{\mathcal{N}=2}$ and the $\mathcal{N}=2$
tensor multiplet $\mathcal{T}^{\mathcal{N}=2}$ used in this paper.

\subsection{Useful tools on type IIB on CY3}

Effective $\mathcal{N}=2$ supersymmetric QFT models in 4d space-time can be
embedded in string compactifications; they are constructed in various
manners; in particular by compactifying type II strings on CY3s or heterotic
string on $K3\times \mathbb{T}^{2}$. These compactifications to 4d $\mathcal{%
N}=2$ low energy theories are related by duality symmetries. By decoupling
massive modes which are of order compactification scale, one can build the
structure of the effective $\mathcal{N}=2$ supersymmetric theory with
decoupled gravity. The massless states following from the CY3
compactification can be organised into $\mathcal{N}=2$ multiplets as follows:%
\newline
$\left( 1\right) $ vectors $\boldsymbol{V}^{\mathcal{N}=2}$; each contains a
complex scalar, a vector and two Weyl fermions.\newline
$\left( 2\right) $ hypermultiplets $\boldsymbol{H}^{\mathcal{N}=2}$; each
contains four real scalars and two Weyl fermions.\newline
$\left( 3\right) $ two other kinds of non standard $\mathcal{N}=2$
multiplets having an antisymmetric tensor $B_{\mu \nu }$; these are: $\left(
a\right) $ the tensor multiplet $\boldsymbol{T}^{\mathcal{N}=2}$ having: 3
scalars, $B_{\mu \nu }$ and 2 Weyl fermions; and $\left( b\right) $ the
vector- tensor $\boldsymbol{R}^{\mathcal{N}=2}$ containing: a scalar, a
vector, $B_{\mu \nu }$ and 2 Weyl fermions.\  \  \  \ 

\emph{Type IIB in 10d}\newline
The \emph{10d} type IIB supergravity multiplet has \emph{128}$+$\emph{128}
on shell degrees of freedom with bosonic sector as follows%
\begin{equation}
\begin{tabular}{lll}
NS-NS & : & $\hat{\phi},$ $\hat{G}_{MN},$ $\hat{B}_{MN}$ \\ 
RR & : & $\hat{C}_{0},$ $\hat{C}_{MN},$ $\hat{C}_{MNPQ}$%
\end{tabular}%
\end{equation}%
In addition to the metric $\hat{G}_{MN}$, we have a real axion $\hat{C}_{0}$
and the dilation $\hat{\phi}=\ln g_{_{IIB}}$ with $g_{_{IIB}}$ the string
coupling. We also have p-forms namely the NS-NS $\hat{B}_{2}$ and the RR p-
form gauge field potentials $\hat{C}_{p}$ together with the corresponding
gauge invariant field strengths 
\begin{equation}
\begin{tabular}{lll}
$\hat{F}_{p+1}$ & $=$ & $d\hat{C}_{p}$ \\ 
$\hat{H}_{3}$ & $=$ & $d\hat{B}_{2}$%
\end{tabular}%
\end{equation}%
This theory has an S- duality symmetry that allows to combine these fields
into $SL\left( 2,Z\right) $ representations; in particular as 
\begin{equation}
\begin{tabular}{lll}
$\tau $ & $=$ & $\hat{C}_{0}-ie^{-\hat{\phi}}$ \\ 
$\hat{G}_{3}$ & $=$ & $\hat{F}_{3}-\tau _{{}}\hat{H}_{3}$ \\ 
$\tilde{F}_{5}$ & $=$ & $\hat{F}_{5}-\frac{1}{2}\hat{C}_{2}\wedge \hat{H}%
_{3}+\frac{1}{2}\hat{B}_{2}\wedge \hat{F}_{3}$%
\end{tabular}%
\end{equation}%
where $\tau $ may be interpreted as the complex structure of 2-torus as in
the embedding of type IIB in F-theory. The 5-form $\tilde{F}_{5}$ gauge
field strength span in 5 of the 10d space time dimensions; it is a self dual
form $\ast \tilde{F}_{5}=\tilde{F}_{5}$ with $\ast $ defined as 
\begin{equation}
\ast \hat{F}^{M_{0}M_{1}...M_{n}}=\frac{1}{n!}\text{ }\varepsilon
^{M_{0}M_{1}...M_{n}M_{n+1}....M_{8}M_{9}}\text{ }\hat{F}%
_{M_{n+1}...M_{9}}^{{}}
\end{equation}%
To avoid confusion between p-form of same rank and also their descendent
after compactification, we shall use the notations $\hat{B}_{2}=\hat{B}%
_{2}^{NS}$ for NS 2-form gauge potential sourced by the elementary string
F1; and $\hat{C}_{2}=\hat{B}_{2}^{RR}$ for the RR 2-form gauge potential
sourced by the solitonic D1 string.

\emph{Type IIB on CY3} \  \  \newline
To descend to 4d space time, we have to factorise the 10d fields as products
of parts; one depending on 4d space time and the other on the internal
coordinates. This is achieved by decomposing the 2- and 4- forms on a
harmonic basis of form of the local CY3 as follows%
\begin{equation}
\begin{tabular}{lll}
$\hat{B}_{2}^{NS}\ $ & $=$ & $B_{2}^{NS}+b_{NS}^{I}\omega _{I}$ \\ 
$\hat{B}_{2}^{RR}$ & $=$ & $B_{2}^{RR}+b_{RR}^{I}\omega _{I}$ \\ 
$\hat{C}_{4}\ $ & $=$ & $C_{4}+A_{2}^{I}\wedge \omega _{I}+V_{1}^{a}\wedge
\alpha _{a}-U_{1a}\beta ^{a}+\varrho _{I}\wedge \tilde{\omega}^{I}$%
\end{tabular}
\label{ed}
\end{equation}%
Here the set $\omega _{I}$, $\alpha _{a},$ $\beta ^{a},$ $\tilde{\omega}^{I}$
stand for a real harmonic basis of p-forms generating the cohomology of the
CY3 obeying amongst others the following useful relations%
\begin{equation}
\dint \nolimits_{CY3}\omega _{I}\wedge \tilde{\omega}^{J}=\delta
_{I}^{J}\qquad ,\qquad \dint \nolimits_{CY3}\mathrm{\alpha }_{a}\wedge 
\mathrm{\beta }^{b}=\delta _{a}^{b}  \label{de}
\end{equation}%
Notice that the 10d self duality condition of $\hat{F}_{5}=d\hat{C}_{4}$
implies that the 2-form $A_{2}^{I}=\frac{1}{2!}A_{\mu \nu }^{I}dx^{\mu
}\wedge dx^{\nu }$ and the scalars $\varrho _{I}$ in eq(\ref{ed}) are
related as $dA_{2}^{I}=\ast d\varrho _{I}$; and so only one of them should
be kept; for our concern we have kept $A_{2}^{I}$. The same feature holds
for the 1-form gauge field potentials $V_{1}^{a}$ and $U_{1a}$; in other
words the decomposition of $\hat{C}_{4}\ $capturing the right number of
degrees of freedom is reduced to%
\begin{equation}
\hat{C}_{4}\ =C_{4}+A_{2}^{I}\wedge \omega _{I}+V_{1}^{a}\wedge \alpha _{a}
\label{xe}
\end{equation}%
Combining altogether, we learn from above decomposition that the 4d space
time spectrum that we obtain, after compactification on CY3, the following
multiplets where only bosonic fields are reported%
\begin{equation}
\begin{tabular}{ll|l|l}
{\small multiplets} &  & \  \  \  \  \  \  \ type IIB/$\mathcal{Z}_{{\small 3}}$ \
\  \  \  & number \\ \hline
{\small gravity } &  & $\  \  \left( \mathcal{G}_{\mu \nu },V_{\mu
}^{0}\right) $ & $\  \ 1$ \\ 
{\small vector } &  & $\  \  \left( V_{\mu }^{a},X^{a},\bar{X}_{a}\right) $ & $%
\  \ h^{2,1}$ \\ 
{\small tensor } &  & $\  \  \left( A_{\mu \nu }^{I},b_{RR}^{I},t^{I},\bar{t}%
^{I}\right) $ & $\  \ h^{1,1}$ \\ 
{\small bi-tensor} &  & $\  \  \left( B_{\mu \nu }^{NS},B_{\mu \nu }^{RR},\xi
,S\right) $ & $\  \ 1$ \\ \hline
\end{tabular}%
\end{equation}%
and where the 4d scalars $\xi $ and $S$ stand respectively the 4d space time
dilaton and 4d axion. \newline
To embed the effective ADJ theory in type IIB string on local CY3, we have
to think about the degrees of freedom of ADJ model to belong to a subsector
of type IIB/CY3 namely 
\begin{equation}
\begin{tabular}{lll}
{\small vector } & : & $\  \  \left( V_{\mu }^{a},X^{a},\bar{X}_{a}\right) $
\\ 
{\small tensor } & : & $\  \  \left( B_{\mu \nu }^{NS},B_{\mu \nu }^{RR},\xi
,S\right) $%
\end{tabular}
\label{ta}
\end{equation}%
with $V_{1}^{a}=V_{\mu }^{a}dx^{\mu }$ standing for 1-form gauge potential
in 4d space time. The fields of (\ref{ta}) are obtained by inverting eqs(\ref%
{ed}) by using properties of the harmonic basis of the homology cycles of
the CY3; they are given by%
\begin{equation}
\begin{tabular}{lll}
$V_{1}^{a}$ & $=$ & $\dint \nolimits_{CY3}\hat{C}_{4}\wedge \mathrm{\beta }%
^{a}$ \\ 
$X^{a}$ & $=$ & $\dint \nolimits_{CY3}\Omega _{3}\wedge \mathrm{\beta }^{a}$%
\end{tabular}%
\end{equation}

\subsection{More on $\mathcal{W}^{\mathcal{N}=2}$ and $\mathcal{T}^{\mathcal{%
N}=2}$ multiplets}

In this appendix, we shed light on those features behind formal similarities
between Maxwell $\mathcal{W}^{\mathcal{N}=2}$ and single tensor $\mathcal{T}%
^{\mathcal{N}=2}$ superfields; this is done by studying their behaviour
under \textrm{the second supersymmetric charge in the fibration (\ref{ZIP});
as well as their link through gauge symmetry}.

\subsubsection{Realising $\mathcal{N}=2$ by two $\mathcal{N}=1$ gauge
superfields}

There are different ways to deal with off shell representations of $\mathcal{%
N}=2$ gauge and tensor multiplets; here we want to focuss on those aspects
shared by their formulation in terms of the $\mathcal{N}=2$ chiral
superfields $\mathcal{W}^{\mathcal{N}=2}$ and $\mathcal{T}^{\mathcal{N}=2%
\text{ }}$ introduced in section 2. These superfields carry $8_{B}+8_{F}$
degrees of freedom; they may be obtained by first considering two $\mathcal{N%
}=1$ superfields having $16_{B}+16_{F}$ degrees of freedom; and then
reducing this number down to $8_{B}+8_{F}$ degrees by imposing appropriate
constraints. The study of the reduction%
\begin{equation}
16_{B}+16_{F}\qquad \rightarrow \qquad 8_{B}+8_{F}  \label{go}
\end{equation}%
for both gauge $\mathcal{W}^{\mathcal{N}=2}$ and tensor $\mathcal{T}^{%
\mathcal{N}=2\text{ }}$ multiplets is interesting in the sense it allows to
get more information on : $\left( i\right) $ the link between the second
supersymmetry and gauge symmetry; and $\left( ii\right) $ the feature behind
similarities between $\mathcal{W}^{\mathcal{N}=2}$ and $\mathcal{T}^{%
\mathcal{N}=2\text{ }}$. Explicitly, we proceed as follows:\newline
First, we use $\mathcal{N}=1$ formalism, where the first supersymmetry is
manifest; to study a simple realisation of $\mathcal{W}^{\mathcal{N}=2}$ and 
$\mathcal{T}^{\mathcal{N}=2}$ engineered from two $\mathcal{N}=1$ hermitian
gauge superfields V$_{1}$ and V$_{2}$; but handled in different manners. 
\newline
After that, we draw the line on how this construction extends to the $%
\mathcal{N}=2$ chiral superspace; this extension is also important for $%
\mathcal{W}^{\mathcal{N}=2}$ and $\mathcal{T}^{\mathcal{N}=2\text{ }}$
because it constitutes the starting point for studying electric-magnetic
duality in $\mathcal{N}=2$ superspace in presence of the Chern-Simons term
given by the coupling 
\begin{equation}
L_{CS}^{\mathcal{N}=2}=\int d^{2}\theta d^{2}\tilde{\theta}\text{ }ig%
\mathcal{W}^{\mathcal{N}=2}\mathcal{T}^{\mathcal{N}=2}+hc
\end{equation}%
Actually, this duality gives another facet on the relationship between $%
\mathcal{W}^{\mathcal{N}=2}$ and $\mathcal{T}^{\mathcal{N}=2}$ as shown by
eq(\ref{wt}); it will not be developed here; but for details see \textrm{%
\cite{1D} }and refs therein.

\  \  \  \  \ 

$\mathcal{N}=2$ \emph{supersymmetry in} $\mathcal{N}=1$ \emph{superspace}%
\newline
In $\mathcal{N}=1$ superspace, the hermitian gauge superfields V$_{1}$ and V$%
_{2}$ describe two representations of $\mathcal{N}=1$ supersymmetry; and
roughly speaking a $\mathcal{N}=2$ multiplet. Altogether, V$_{1}$ and V$_{2}$
carry $16_{B}+16_{F}$ off shell degrees of freedom; half of them coming from
V$_{1}$ and the other half from V$_{2}$. With these two superfields, the
second supersymmetry is generated by the transformations%
\begin{equation}
\begin{tabular}{lll}
$\tilde{\delta}V_{1}$ & $=$ & $\frac{-i}{\sqrt{2}}\left( \epsilon D+\bar{%
\epsilon}\bar{D}\right) V_{2}$ \\ 
$\tilde{\delta}V_{2}$ & $=$ & $i\sqrt{2}\left( \epsilon D+\bar{\epsilon}_{{}}%
\bar{D}\right) V_{1}$%
\end{tabular}%
\end{equation}%
with supersymmetric parameter $\epsilon =\delta \tilde{\theta}$. These $%
16_{B}+16_{F}$ degrees can be reduced down to $8_{B}+8_{F}$ by imposing
constraint eqs on V$_{1}$ and V$_{2}$; it happens that this can be done in
two different manners; one leading to $\left( \boldsymbol{X},\boldsymbol{W}%
_{\alpha }\right) $, the $\mathcal{N}=1$ superfields representation of $%
\mathcal{W}^{\mathcal{N}=2}$; and the other to the $\left( \boldsymbol{\Phi }%
,\boldsymbol{L}\right) $ realisation of $\mathcal{T}^{\mathcal{N}=2}$; or up
to a gauge symmetry, to $\left( \boldsymbol{Y},\boldsymbol{\chi }_{\alpha },%
\boldsymbol{\Phi }\right) $ with $\boldsymbol{L}=D^{\alpha }\boldsymbol{\chi 
}_{\alpha }+\bar{D}_{\dot{\alpha}}\boldsymbol{\bar{\chi}}^{\dot{\alpha}}$
and $\boldsymbol{Y}$ as in (\ref{Y}).

\  \ 

\emph{From }$\left( \emph{V}_{1}\emph{,V}_{2}\right) $\emph{\ to} $\mathcal{W%
}^{\mathcal{N}=2}$\newline
In this realisation, the $\mathcal{N}=1$ hermitian superfields V$_{1}$ and V$%
_{2}$ are interpreted as gauge superfield potentials obeying the following U$%
\left( 1\right) $ gauge transformations%
\begin{equation}
\begin{tabular}{lllll}
$V_{1}^{\prime }$ &  & $\equiv $ &  & $V_{1}+\boldsymbol{\Lambda }_{l}$ \\ 
$V_{2}^{\prime }$ &  & $\equiv $ &  & $V_{2}+\left( \boldsymbol{\Lambda }%
_{c}+\boldsymbol{\bar{\Lambda}}_{c}\right) $%
\end{tabular}
\label{C1}
\end{equation}%
They involve two $\mathcal{N}=1$ superfield gauge parameters; the hermitian $%
\boldsymbol{\Lambda }_{l}=\boldsymbol{\bar{\Lambda}}_{l}$ and the chiral $%
\boldsymbol{\Lambda }_{c}$ solving the superspace conditions 
\begin{equation}
\begin{tabular}{lllll}
$D^{2}\Lambda _{l}$ & $=$ & $0$ & $=$ & $\bar{D}^{2}\Lambda _{l}$ \\ 
$\bar{D}_{\dot{\alpha}}\Lambda _{c}$ & $=$ & $0$ & $=$ & $D_{\alpha }\bar{%
\Lambda}_{c}$%
\end{tabular}%
\end{equation}%
Each one of these superparameters reduces by 4 the number of degrees of
freedom in the corresponding gauge superfield. The $\mathcal{N}=2$ Maxwell
chiral superfield strength $\mathcal{W}^{\mathcal{N}=2}\equiv \left( 
\boldsymbol{X},\boldsymbol{W}_{\alpha }\right) $ is related to $\left(
V_{1},V_{2}\right) $ through the following gauge invariant quantities%
\begin{equation}
\boldsymbol{X}=\frac{1}{2}\bar{D}^{2}V_{1}\qquad ,\qquad \boldsymbol{W}%
_{\alpha }=-\frac{1}{4}\bar{D}^{2}D_{\alpha }V_{2}  \label{rt}
\end{equation}%
they can be combined into a $\mathcal{N}=2$ chiral superfield strength as in
(\ref{22}) namely%
\begin{equation*}
\mathcal{W}^{\mathcal{N}=2}=\boldsymbol{X}+i\sqrt{2}\text{ }\tilde{\theta}%
^{\alpha }\boldsymbol{W}_{\alpha }-\tilde{\theta}^{2}\left( \frac{1}{4}\bar{D%
}^{2}\boldsymbol{\bar{X}}_{{}}\right)
\end{equation*}%
which by substituting can be also expressed like $\mathcal{W}^{\mathcal{N}%
=2}=\frac{1}{4}\bar{D}^{2}\Gamma $ with 
\begin{equation}
\Gamma =2V_{1}-i\sqrt{2}\tilde{\theta}^{\alpha }\left( D_{\alpha
}V_{2}\right) -\tilde{\theta}^{2}\left( D^{2}V_{1}\right)
\end{equation}%
Notice that in analogy with the Wess-Zumino gauge commonly used for $V_{2}$,
the vector superfield $V_{1}$ in (\ref{rt}) has also a Wess-Zumino like
expansion leading, after substituting in $\boldsymbol{X}=\frac{1}{2}\bar{D}%
^{2}V_{1},$ to the following $\theta $- expansion%
\begin{equation}
\boldsymbol{X}=X+\sqrt{2}\theta \psi _{1}+\theta ^{2}\bar{\theta}^{2}F_{X}
\label{X1}
\end{equation}%
with $X$ standing for the complex scalar of the $\mathcal{N}=2$ gauge
multiplet and $\psi _{\alpha 1}$ for one of the two gauginos; the other
gaugino $\psi _{\alpha 2}$ comes from $\boldsymbol{W}_{\alpha }$. However,
the non propagating complex auxiliary field $F_{X}$ is realised here as $%
F_{X}=-d_{1}-i\partial _{\mu }\upsilon _{1}^{\mu }$ with imaginary part $%
\func{Im}F_{X}$ given by the topological quantity 
\begin{equation}
\partial _{\mu }\upsilon _{1}^{\mu }=\frac{i}{4!}\varepsilon ^{\mu \nu \rho
\sigma }\mathcal{F}_{[\mu \nu \rho \sigma ]}  \label{X2}
\end{equation}%
with $\mathcal{F}_{[\mu \nu \rho \sigma ]}$ interpreted as the field
strength of a 3-form gauge potential $\mathcal{A}_{[\nu \rho \sigma ]}$
describing a non-propagating component field. Being completely
antisymmetric, this 4-tensor $\mathcal{F}_{[\mu \nu \rho \sigma ]}$ can be
realised like $\frac{1}{\kappa }\varepsilon _{\mu \nu \rho \sigma }$ where $%
\frac{1}{\kappa }$ is precisely the deformation term appearing in the second
line of (\ref{22}).

\subsubsection{From\emph{\ }$\left( \emph{V}_{1}\emph{,V}_{2}\right) $ to
tensor multiplet $\mathcal{T}^{\mathcal{N}=2}$}

First recall that in this paper, we have considered two kinds of
realisations of the tensor multiplet $\mathcal{T}^{\mathcal{N}=2}$: $\left(
i\right) $ a short representation where $\mathcal{T}^{\mathcal{N}=2}\equiv
\left( \boldsymbol{L},\boldsymbol{\Phi }\right) $ having $8_{B}+8_{F}$
degrees of freedom; and $\left( ii\right) $ a long representation $\mathcal{T%
}^{\mathcal{N}=2}\equiv \left( \boldsymbol{Y},\boldsymbol{\chi }_{\alpha },%
\boldsymbol{\Phi }\right) $ involving $16_{B}+16_{F}$ degrees of freedom
obeying the symmetry (\ref{gh}). The first one is obtained from the second
by gauging away half of the degrees of freedom as in eq(\ref{go}). In fact
both short and long representation of $\mathcal{T}^{\mathcal{N}=2}$ may be
imagined as solutions of the following linear constraint relations%
\begin{equation}
\begin{tabular}{lllll}
$\bar{D}^{2}U_{1}$ & $=$ & $D^{2}U_{1}$ & $=$ & $0$ \\ 
$\bar{D}^{2}D_{\alpha }U_{2}$ & $=$ & $D^{2}\bar{D}_{\dot{\alpha}}U_{2}$ & $%
= $ & $0$%
\end{tabular}
\label{C2}
\end{equation}%
where we have used the notation $\left( U_{1},U_{2}\right) $ to avoid
confusion with the hermitian multiplets $\left( V_{1},V_{2}\right) $ used in
eqs(\ref{rt}). Notice that eqs(\ref{C2}) may thought of as the complement of
eqs(\ref{rt}) in the space parameterized by $\left( V_{1},V_{2}\right) $.%
\newline
A first type of solution of these constraint relations is easily identified
by remembering that the $\mathcal{N}=1$ linear multiplet $\boldsymbol{L}$
satisfies also the conditions $D^{2}\boldsymbol{L}=\bar{D}^{2}\boldsymbol{L}%
=0$ exactly as $U_{1}$. Moreover because of the chirality properties $\bar{D}%
_{\dot{\alpha}}\boldsymbol{\Phi }=D_{\alpha }\boldsymbol{\Phi }=0$ as well
as the relations $\left[ D_{\alpha },\partial _{\mu }\right] =\left[ \bar{D}%
_{\dot{\alpha}},\partial _{\mu }\right] =0$; it follows that $\left(
U_{1},U_{2}\right) $ are nothing but 
\begin{equation}
\begin{tabular}{ll}
$U_{1}=$ & $\boldsymbol{L}$ \\ 
$U_{2}=$ & $\boldsymbol{\Phi }+\boldsymbol{\bar{\Phi}}$%
\end{tabular}%
\end{equation}%
where we have dropped out the spurious superfield $\boldsymbol{Y}$. \newline
The second type of solution of (\ref{C2}) is given by $\mathcal{T}^{\mathcal{%
N}=2}\equiv \left( \boldsymbol{Y},\boldsymbol{\chi }_{\alpha },\boldsymbol{%
\Phi }\right) $; it is convenient for the use of the $\mathcal{N}=2$ chiral
superspace and reads, roughly speaking, in terms of the four $\mathcal{N}=1$
chiral superfields as follows%
\begin{equation}
\begin{tabular}{ll}
$U_{1}=$ & $D^{\alpha }\boldsymbol{\chi }_{\alpha }+\bar{D}_{\dot{\alpha}}%
\boldsymbol{\bar{\chi}}^{\dot{\alpha}}$ \\ 
$U_{2}=$ & $\left( \boldsymbol{\Phi }+\boldsymbol{\bar{\Phi}}\right) +\frac{i%
}{2}\left( D^{2}\boldsymbol{Y}-\bar{D}^{2}\boldsymbol{\bar{Y}}\right) $%
\end{tabular}%
\end{equation}%
with the gauge symmetry property%
\begin{equation}
\begin{tabular}{lll}
$\boldsymbol{\chi }_{\alpha }^{\prime }$ & $=$ & $\boldsymbol{\chi }_{\alpha
}+\frac{i}{4}\bar{D}^{2}D_{\alpha }\Omega $ \\ 
$\boldsymbol{Y}^{\prime }$ & $=$ & $\boldsymbol{Y}-\frac{1}{2}\bar{D}%
^{2}\Upsilon $ \\ 
$\boldsymbol{\Phi }^{\prime }$ & $=$ & $\boldsymbol{\Phi }$%
\end{tabular}
\label{gh}
\end{equation}%
where $\Omega $ and $\Upsilon $ are two $\mathcal{N}=1$ hermitian gauge
superfield parameters. Notice that the implementation of the superfield $%
\boldsymbol{Y}$ is required by off shell closure of $\mathcal{N}=2$
supersymmetry as in eq(2.8). Notice also the three following features:%
\newline
$\left( i\right) $ first in the gauge where $\boldsymbol{Y}^{\prime }=0$,
the chiral superfield $\boldsymbol{Y}$ is given by $\frac{1}{2}\bar{D}%
^{2}\Upsilon $; by comparing this expression with the first relation of (\ref%
{rt}), one learns that $\boldsymbol{Y}$ has same form as the chiral
superfield $\boldsymbol{X}=\frac{1}{2}\bar{D}^{2}V_{1}$. Obviously $%
\boldsymbol{X}$ and $\boldsymbol{Y}$ are different things; the first one
carries propagating physical degrees of freedom; while the second is a pure
auxiliary superfield with no physical field.\newline
$\left( ii\right) $ second alike for the relations (\ref{X1}-\ref{X2})
satisfied by the $\boldsymbol{X}$ superfield and allowing adjunction of a
topological term to the lagrangian density implemented in superspace
formulation by the $\frac{\tilde{\theta}^{2}}{2\kappa }$ deformation in the
second line of (\ref{22}), one has as well a quite similar feature for the
superfield $\boldsymbol{Y}$. Indeed, following \textrm{\cite{1D}, there is a
gauge where }$\boldsymbol{Y}$ reads as 
\begin{equation}
\boldsymbol{Y}=\frac{i}{4!}\theta ^{2}\varepsilon ^{\mu \nu \rho \sigma
}C_{\mu \nu \rho \sigma }  \label{X3}
\end{equation}%
with $C_{\mu \nu \rho \sigma }$ a 4-form field with no propagating degree of
freedom.\newline
$\left( iii\right) $ third if denoting by%
\begin{equation}
\begin{tabular}{lll}
$\hat{w}_{\alpha }$ & $=$ & $-\frac{1}{4}\bar{D}^{2}D_{\alpha }\Omega $ \\ 
$\hat{x}$ & $=$ & $\frac{1}{2}\bar{D}^{2}\Upsilon $%
\end{tabular}%
\end{equation}%
then the gauge transformations (\ref{gh}) read as 
\begin{equation}
\begin{tabular}{lll}
$\boldsymbol{\chi }_{\alpha }^{\prime }$ & $=$ & $\boldsymbol{\chi }_{\alpha
}-i\hat{w}_{\alpha }$ \\ 
$\boldsymbol{Y}^{\prime }$ & $=$ & $\boldsymbol{Y}-\hat{x}$ \\ 
$\boldsymbol{\Phi }^{\prime }$ & $=$ & $\boldsymbol{\Phi }$%
\end{tabular}%
\end{equation}%
Putting this change into 
\begin{equation*}
\mathcal{T}^{\mathcal{N}=2}=\boldsymbol{Y}+\sqrt{2}\tilde{\theta}^{\alpha }%
\boldsymbol{\chi }_{\alpha }-\tilde{\theta}^{2}\left( \frac{i}{2}\boldsymbol{%
\Phi }+\frac{1}{4}\bar{D}^{2}\boldsymbol{\bar{Y}}\right)
\end{equation*}%
we find that it transforms like 
\begin{equation}
\mathcal{T}^{^{\prime }\mathcal{N}=2}=\mathcal{T}^{\mathcal{N}=2}-\hat{w}
\label{tw}
\end{equation}%
with $\mathcal{N}=2$ superfield gauge parameter%
\begin{equation}
\hat{w}=\hat{x}+i\sqrt{2}\tilde{\theta}^{\alpha }\hat{w}_{\alpha }-\frac{1}{4%
}\tilde{\theta}^{2}\bar{D}^{2}\overline{\hat{x}}  \label{ww}
\end{equation}%
having the same structure as the Maxwell superfield $\mathcal{W}^{\mathcal{N}%
=2}$. From this view, it follows that a single-tensor superfield $\mathcal{T}%
^{\mathcal{N}=2}$ is a chiral superfield%
\begin{equation}
\mathcal{Z}=Z\left( y,\theta \right) +\sqrt{2}\tilde{\theta}^{\alpha
}\Upsilon _{\alpha }\left( y,\theta \right) -\tilde{\theta}^{2}F\left(
y,\theta \right) \left[ \frac{i}{2}\Phi _{Z}\left( y,\theta \right) +\frac{1%
}{4}\bar{D}^{2}\bar{Z}\left( y,\theta \right) \right]
\end{equation}%
obeying the gauge symmetry (\ref{tw}) with superfield parameter as in (\ref%
{ww}).

\subsubsection{Comment on gauge symmetry}

Here we comment on the gauge symmetry of the field action term $\mathcal{S}%
\left( Y\right) =\int d^{4}x\mathcal{L}\left( Y\right) $ with lagrangian
density $\mathcal{L}\left( Y\right) \equiv \mathcal{L}$ given by 
\begin{equation}
\mathcal{L}=\eta \int d^{2}\theta _{{}}\mathbf{Y}+hc  \label{yy}
\end{equation}%
and coupling constant a pure imaginary number $\eta =i\nu $. This field
action $\mathcal{S}\left( Y\right) $ is an extra term that has been used in
this paper to induce a deformation of the ADJ constraint $\sum_{a}\frac{g_{a}%
}{\kappa _{a}}=0$ into $\nu +\sum_{a}\frac{g_{a}}{\kappa _{a}}=0$; it lifts
the singularity of the particular equation $\frac{g_{1}}{\kappa _{1}}=0$;
but breaks $\mathcal{N}=2$ supersymmetry explicitly. \newline
Gauge invariance of $\mathcal{S}\left( Y\right) $ can be studied either
directly by using the superspace method taking into account that the
superfield $\mathbf{Y}$ is a chiral multiplet; or more explicitly by working
with component fields. In the first way, the gauge transformation of $%
\mathbf{Y}$ can be learnt from (\ref{gh}); it reads as $\mathbf{Y}^{\prime }=%
\boldsymbol{Y}+\delta _{gauge}\mathbf{Y}$ with $\delta _{gauge}\mathbf{Y=}-%
\frac{1}{2}\bar{D}^{2}\Upsilon $; where $\Upsilon $ is an arbitrary $%
\mathcal{N}=1$ real superfield. To shed light on this gauge transformation
and its effect on the action $\mathcal{S}\left( Y\right) $; we will use the
component field language to first build the explicit $\theta $-expansions of 
$\Upsilon $ and $\bar{D}^{2}\Upsilon $; then turn back to study gauge
symmetry of (\ref{yy}).

\  \  \  \  \ 

$\bullet $ \emph{component field analysis}\  \newline
We begin by describing the $\theta $-expansion of the superfield $Y$ which
can be expressed into two manners: $\left( i\right) $ either by using the
complex chiral superspace coordinate basis $Z_{c}=\left( x-i\theta \sigma 
\bar{\theta},\theta \right) $ where $Y$ is expanded as $y+\sqrt{2}\theta
\Psi +\theta ^{2}F$; or $\left( ii\right) $ by working in the real
superspace coordinate basis $Z_{R}=\left( x^{\mu },\theta ^{a},\bar{\theta}_{%
\dot{a}}\right) $ where the superfield Y has also an explicit $\bar{\theta}$
dependence as shown below\textrm{\footnote{%
Our notations are as in Wess-Bagger's book: supersymmetry and Supergravity 
\cite{WB}.}} 
\begin{equation}
Y=y-i\theta \sigma ^{\mu }\bar{\theta}\partial _{\mu }y+\frac{1}{4}\theta
^{2}\bar{\theta}^{2}\square y+\sqrt{2}\theta \Psi +\frac{i}{\sqrt{2}}\theta
^{2}\partial _{\mu }\Psi \sigma ^{\mu }\bar{\theta}+\theta ^{2}F_{Y}
\label{ya}
\end{equation}%
From this expansion, we learn 
\begin{equation}
\mathcal{L}=i\nu \left( F_{Y}-\bar{F}_{Y}\right)
\end{equation}%
In the gauge transformation $\delta _{gauge}\mathbf{Y=-}\frac{1}{2}\bar{D}%
^{2}\Upsilon $, the real superfield parameter $\Upsilon $ has a $\theta $-
expansion involving as usual several component field parameters which, for
later use can be expressed like 
\begin{equation}
\Upsilon =\gamma -\frac{1}{2}\theta ^{2}\bar{M}-\frac{1}{2}\bar{\theta}%
^{2}M-\theta \sigma ^{\mu }\bar{\theta}w_{\mu }+\theta ^{2}\bar{\theta}%
^{2}\left( d_{\Upsilon }+\frac{1}{4}\square \gamma \right) +fermionic
\label{gam}
\end{equation}%
where "\emph{fermionic}" stands for those monomials with odd powers in
Grassman variables involving fermionic fields. The components $\gamma $, $%
d_{\Upsilon }$ are real scalar components; the complex scalars $M$, $\bar{M}$
are related by complex conjugation; and the real 4-vector $w_{\mu }$ is
parameterised like $w_{\mu }=\frac{1}{3!}\varepsilon _{\mu \nu \rho \sigma
}A^{\left[ \nu \rho \sigma \right] }$ with $A^{\left[ \nu \rho \sigma \right]
}$ a completely antisymmetric real rank 3- tensor. This dual parametrisation
of $w_{\mu }$ is dictated by an underlying arbitrariness in dealing with the
superfield $\Upsilon $; by shifting the real superparameter $\Upsilon $ by a
hermitian superfield $\Gamma $ like $\Upsilon ^{\prime }=\Upsilon +\Gamma $,
the quantity $\bar{D}^{2}\Upsilon $ remains invariant provided $\Gamma $
satisfying the constraints 
\begin{equation}
\bar{D}^{2}\Gamma =D^{2}\Gamma =0\qquad ,\qquad \Gamma ^{\dagger }=\Gamma
\label{ar}
\end{equation}%
But these conditions are same as the ones defining linear multiplet $%
\boldsymbol{L}$ introduced in section 2 and which we recall here for
comparison $\bar{D}^{2}\boldsymbol{L}=D^{2}\boldsymbol{L}=0$ with $%
\boldsymbol{L}^{\dagger }=\boldsymbol{L}$; the real superfield $\Gamma $ has
therefore a similar $\theta $- expansion as eq(\ref{216}); and so it can be
expressed like%
\begin{equation}
\Gamma =\gamma +\theta \sigma ^{\mu }\bar{\theta}\varepsilon _{\mu \nu \rho
\sigma }\partial ^{\nu }\omega ^{\left[ \rho \sigma \right] }-\frac{1}{4}%
\theta ^{2}\bar{\theta}^{2}\square \gamma +fermionic
\end{equation}%
showing that the term $\varepsilon _{\mu \nu \rho \sigma }\partial ^{\nu
}\omega ^{\rho \sigma }$ is nothing but a gauge transformation of $w_{\mu }=%
\frac{1}{3!}\varepsilon _{\mu \nu \rho \sigma }A^{\left[ \nu \rho \sigma %
\right] }$ with gauge parameter given by the rank 2- antisymmetric $\omega
^{\rho \sigma }=-\omega ^{\sigma \rho }$. Taking advantage of the
arbitrariness (\ref{ar}), one can make convenient choices; in particular we
can put the real superfield $\Upsilon $ into the following form; see also 
\textrm{\cite{1D},} 
\begin{equation}
\Upsilon =-\frac{1}{2}\theta ^{2}\bar{M}-\frac{1}{2}\bar{\theta}^{2}M-\theta
\sigma ^{\mu }\bar{\theta}w_{\mu }+\frac{1}{\sqrt{2}}\theta ^{2}\bar{\theta}%
\bar{\lambda}+\frac{1}{\sqrt{2}}\bar{\theta}^{2}\theta \lambda +\frac{1}{2}%
\theta ^{2}\bar{\theta}^{2}d_{\Upsilon }  \label{am}
\end{equation}%
With this choice; the initial gauge symmetry (\ref{gh}) gets now reduced to
gauge symmetry of the 4- vector $w_{\mu }=\frac{1}{3!}\varepsilon _{\mu \nu
\rho \sigma }A^{\left[ \nu \rho \sigma \right] }$ namely 
\begin{equation}
w_{\mu }\qquad \rightarrow \qquad w_{\mu }+\varepsilon _{\mu \nu \rho \sigma
}\partial ^{\nu }\omega ^{\rho \sigma }  \label{gs}
\end{equation}

$\bullet $ \emph{more on gauge invariance}\newline
From the gauge fixed expression (\ref{am}), we can make two useful
computations namely the explicit expressions of $\frac{1}{2}\bar{D}%
^{2}\Upsilon $ and the integral over Grassman variables $i\nu \int
d^{2}\theta $ $\boldsymbol{Y}+hc$. For the $\theta $- expansion of $\frac{1}{%
2}\bar{D}^{2}\Upsilon $; we find%
\begin{equation}
\begin{tabular}{lll}
$\frac{1}{2}\bar{D}^{2}\Upsilon $ & $=$ & $M+\sqrt{2}\theta \lambda -\theta
^{2}\left[ d_{\Upsilon }+i\partial _{\mu }w^{\mu }\right] +i\theta \sigma
^{\nu }\bar{\theta}\partial _{\nu }M$ \\ 
&  & $-\frac{i}{\sqrt{2}}\theta ^{2}\partial _{\nu }\lambda \sigma ^{\nu }%
\bar{\theta}+\frac{1}{4}\theta ^{2}\bar{\theta}^{2}\square M$%
\end{tabular}
\label{ma}
\end{equation}%
where 
\begin{equation}
\begin{tabular}{lll}
$\partial _{\mu }w^{\mu }$ & $=$ & $\frac{1}{3!}\varepsilon ^{\mu \nu \rho
\sigma }\partial _{\mu }A_{\left[ \nu \rho \sigma \right] }$ \\ 
& $=$ & $\frac{1}{4!}\varepsilon ^{\mu \nu \rho \sigma }F_{[\mu \nu \rho
\sigma ]}\equiv \Delta $%
\end{tabular}%
\end{equation}%
has an interpretation in term of rank 4- antisymmetric field strength $%
F_{[\mu \nu \rho \sigma ]}$ of the 3- form potential field $A_{\left[ \nu
\rho \sigma \right] }$. Under the residual gauge (\ref{gs}), the term $%
\partial _{\mu }w^{\mu }$ is therefore manifestly gauge invariant; this
property can be explicitly checked on the transformation%
\begin{equation}
\partial _{\mu }w^{\mu }\qquad \rightarrow \qquad \partial _{\mu }w^{\mu
}+\varepsilon ^{\mu \nu \rho \sigma }\partial _{\mu }\partial _{\nu }\omega
_{\rho \sigma }  \label{du}
\end{equation}%
where the extra term $\varepsilon ^{\mu \nu \rho \sigma }\partial _{\mu
}\partial _{\nu }\omega _{\rho \sigma }$ vanishes identically due to a
tensor calculus feature. \newline
Comparing (\ref{ya}) and (\ref{ma}), one remarks, that by using a
Wess-Zumino-like gauge, $\mathbf{Y}$ can be expressed as $\boldsymbol{Y}_{%
{\small gauged}}=i\theta ^{2}\partial _{\mu }w^{\mu }$. Moreover, using this
gauge fixed expression, the field action $\mathcal{S}=\int d^{4}x\mathcal{L}$
with the lagrangian density $\mathcal{L}$\ as in eq(\ref{yy}) reads
therefore like, 
\begin{equation}
\mathcal{S}=-2\nu \int d^{4}x\text{ }\partial _{\mu }w^{\mu }  \label{sg}
\end{equation}%
and, because of (\ref{du}), is manifestly gauge invariant $\delta _{gauge}%
\mathcal{S}=0$. \newline
In the end of this comment, we would like to add that even if thinking of
the scalar $\partial ^{\mu }w_{\mu }=\Delta $ as the dual of a generic rank
4- tensor gauge potential like $\Delta =\frac{1}{4!}\varepsilon ^{\mu \nu
\rho \sigma }A_{[\mu \nu \rho \sigma ]}$ where the completely antisymmetric $%
A_{[\mu \nu \rho \sigma ]}$ obeys the gauge transformation $\delta A_{\left[
\mu \nu \rho \sigma \right] }=\partial _{\lbrack \mu }\Lambda _{\nu \rho
\sigma ]}$, the variation of the quantity $\frac{1}{4!}\varepsilon ^{\mu \nu
\rho \sigma }A_{[\mu \nu \rho \sigma ]}$ behaves as a divergence term $\frac{%
1}{3!}\partial _{\mu }\left( \varepsilon ^{\mu \nu \rho \sigma }\Lambda
_{\nu \rho \sigma }\right) $; and, up on ignoring boundary effects, this
variation does not contribute at the level of the action $\mathcal{S}=-2\nu
\int d^{4}x$ $\frac{1}{4!}\varepsilon ^{\mu \nu \rho \sigma }A_{[\mu \nu
\rho \sigma ]}$. However, in this way of doing, the field strength $F_{[\mu
\nu \rho \sigma \tau ]}\equiv F_{\left( 5\right) }$, associated to the
potential field $A_{[\mu \nu \rho \sigma ]}\equiv A_{\left( 4\right) }$,
would be a rank 5- antisymmetric tensor field which is invariant under the
gauge transformation $A_{\left( 4\right) }\rightarrow A_{\left( 4\right)
}+d\Lambda _{\left( 3\right) }$; but because of space time dimension
constraint, the rank 5- tensor should be equal to zero, $F_{\left( 5\right)
}=dA_{\left( 4\right) }=0$; this means that the 4-form $A_{\left( 4\right) }$
is a pure gauge potential without curvature which may be thought of as $%
A_{\left( 4\right) }=dA_{\left( 3\right) }$; and then the gauge parameter $%
\Lambda _{\left( 3\right) }$ as just a shift of the origin of $A_{\left(
3\right) }$.

\begin{acknowledgement}
Saidi would like to thank the ICTP- Senior Associate programme for
supporting his stay at the International Centre for Theoretical Physics,
Trieste Italy; where this work has been revised.
\end{acknowledgement}


\begin{thebibliography}{99}
\bibitem{1A} E. Cremmer et al. Nucl.Phys. B250(1985)385,

\bibitem{2A} S.Ferrara , P. van Nieuwenhuizen Phys.Let. 127B(1983)70,

\bibitem{3A} Yu. M. Zinoviev, Yad. Fiz. 46 (1987) 1240,

\bibitem{4A} Ignatios Antoniadis, H. Partouche, T.R. Taylor, Phys.Lett. B372
(1996) 83-87,

\bibitem{5A} S. Ferrara, L. Girardello and M. Porrati, Phys. Lett. B 366
(1996) 155. [arXiv:hep-th/9510074],

\bibitem{6A} S. Ferrara, L. Girardello, M. Porrati, Phys.Lett. B376 (1996)
275-281,

\bibitem{60A} S. Cecotti, L. Girardello, and M. Porrati, Nucl. Phys. B268
(1986) 295,

\bibitem{7A} E. Cremmer, C. Kounnas, Antoine Van Proeyen, J.P. Derendinger,
S. Ferrara, B. de Wit, L. Girardello, Nucl.Phys. B250 (1985) 385,

\bibitem{8A} Hiroshi Itoyama, Kazunobu Maruyoshi, Makoto Sakaguchi,
Nucl.Phys. B794 (2008) 216-230, e-Print: arXiv:0709.3166,

\bibitem{9A} M. De Roo, P. Wagemans, Physics Letters B, 177, 352--356,
(1986),

\bibitem{10A} E. Kiritsis, C. Kounnas, Nucl.Phys. B503 (1997) 117-156,
arXiv:hep-th/9703059,

\bibitem{11A} Tomasz R. Taylor, Cumrun Vafa, Physics Letters B, 474,
130--137, (2000),

\bibitem{12A} B. de Wit, P. G. Lauwers, and A. Van Proeyen, Nucl. Phys. B
255 (1985) 569,

\bibitem{1AB} J. Bagger and J. Wess, Phys. Lett. B138, (1984) 105,

\bibitem{1BA} A. El Hassouni - E. G. Oudrhiri-Safiani, E. H. Saidi - JMP 28,
(1987) 2457,

\bibitem{1AC} J. Hughes, J. Liu and J. Polchinski, Supermembranes, Phys.
Lett. B180 (1986) 370,

\bibitem{1AD} H. Partouche and B. Pioline, Nucl. Phys. Proc. Suppl. 56B
(1997) 322, hep-th/9702115,

\bibitem{13A} P. Fayet, Fermi-Bose hypersymmetry, Nucl. Phys. B 113 (1976)
135,

\bibitem{1AX} Laura Andrianopoli, Riccardo D'Auria, Sergio Ferrara, Mario
Trigiante, \emph{Observations on the Partial Breaking of }$\mathcal{N}=2$%
\emph{\ Rigid Supersymmetry},  arXiv:1501.07842,

\bibitem{1AY} Laura Andrianopoli, Patrick Concha, Riccardo D'Auria, Evelyn
Rodriguez, Mario Trigiante, \emph{Observations on BI from }$\mathcal{N}=2$%
\emph{\ Supergravity and the General Ward Identity},  arXiv:1508.01474

\bibitem{1AE} P. Fre, L. Girardello, I. Pesando and M. Trigiante, Nucl.
Phys. B493 (1997)231, hep-th/9607032,

\bibitem{1B} Hans Peter Nilles; the strings connection, Euro. Phys. J. C 74
(2014) 2712; in Supersymmetry after the Higgs Discovery; Editors, I
Antoniadis, D. Ghilencia,

\bibitem{2B} H. Itoyama, Nobuhito Maru, Int. J. Mod. Phys. A27 (2012)
1250159, arXiv:1109.2276 [hep-ph],

\bibitem{3B} Benoit L\'{e}geret, Claudio A. Scrucca and Paul Smyth,
Phys.Lett. B722 (2013) 372-377 arXiv:1211.7364 [hep-th],

\bibitem{4B} I. Antoniadis, E. Dudas, D. M. Ghilencea, P. Tziveloglou, Nucl.
Phys. B841 (2010) 157-177. [arXiv:1006.1662 [hep-ph]],

\bibitem{1C} J.-C. Jacot and C. A. Scrucca, Nucl. Phys. B 840 (2010) 67
[arXiv:1005.2523],

\bibitem{2C} I. Antoniadis, J.-P. Derendinger, T. Maillard,\emph{\ }Nucl.
Phys. B 808 (2009) 53\emph{\ }arXiv:0804.1738 [hep-th],

\bibitem{3C} J. Bagger and A. Galperin, Phys. Lett. B336 (1994) 25,

\bibitem{4C} J. Bagger , A. Galperin,Phys. Rev. D 55 (1997) 1091
[arXiv:hep-th/9608177],

\bibitem{5C} Jonathan Bagger, Chi Xiong, N=2 nonlinear sigma models in $%
\mathcal{N}=1$ superspace: Four and five dimensions, e-Print: hep-th/0601165,

\bibitem{6C} E.A. Ivanov and B.M. Zupnik, Modified $\mathcal{N}=2$
supersymmetry and Fayet-Iliopoulos terms, Phys.Atom.Nucl., 62:1043--1055,
1999, arXiv:hep-th/9710236,

\bibitem{7C} H. Partouche and B. Pioline, Partial spontaneous breaking of
global supersymmetry, Nucl.Phys.Proc.Suppl., 56B:322--327, 1997,

\bibitem{1D} N. Ambrosetti, I. Antoniadis, J.-P. Derendinger, and P.
Tziveloglou, Nucl.Phys., B835:75-109, 2010, arXiv:0911.5212,

\bibitem{2D} I. Antoniadis, J. -P. Derendinger and J. -C. Jacot, Nucl. Phys.
B 863 (2012) 471 [arXiv:1204.2141],

\bibitem{3D} I. Antoniadis and M. Buican, Goldstinos,\emph{\ }JHEP 1104
(2011) 101 [arXiv:1005.3012],

\bibitem{1G} Steven B. Giddings, Shamit Kachru, Joseph Polchinski,
hep-th/0105097,

\bibitem{1E} U. Lindstrom and M. Rocek, Nucl. Phys. B 222 (1983) 285,

\bibitem{2E} N. J. Hitchin, A. Karlhede, U. Lindstrom and M. Rocek, Commun.
Math. Phys. 108 (1987) 535,

\bibitem{3E} M. Grana, J. Louis and D. Waldram, JHEP 0601, 008 (2006)
[arXiv:hep-th/0505264],

\bibitem{4E} A. Galperin, E. Ivanov, S. Kalitsyn, V. Ogievetsky and E.
Sokatchev, Class. Quant.Grav. 1 (1984) 469,

\bibitem{5E} El Hassan Saidi, Nucl.Phys.B803:323-362,2008, arXiv:0806.3207,

\bibitem{CY3} P. Candelas and X. de la Ossa, \textquotedblleft Moduli Space
of Calabi-Yau Manifolds,\textquotedblright \ Nucl. Phys.B355, 455 (1991),

\bibitem{5EA} Malika Ait Benhaddou, El Hassan Saidi, Physics Letters
B575(2003)100-110, arXiv:hep-th/0307103,

\bibitem{5EB} M. Ait Ben Haddou, A. Belhaj, E.H. Saidi, Nucl.Phys. B674
(2003) 593-614, arXiv:hep-th/0307244,

\bibitem{5EC} R. Ahl Laamara, M. Ait Ben Haddou, A Belhaj, L.B Drissi, E.H
Saidi, Nucl.Phys. B702 (2004) 163-188, arXiv:hep-th/0405222,

\bibitem{2G} G. Aldazabal, Badagnani, L. E. Ibanez, A. M. Uranga, JHEP 9906
(1999) 031 [arXiv:hep-th/9904071],

\bibitem{3G} A. M. Uranga, Nucl.Phys.B598:225-246, (2001),
arXiv:hep-th/0011048,

\bibitem{GVW} S. Gukov, C. Vafa and E. Witten, Nucl. Phys. B 584 (2000) 69
[Erratum-ibid. B608 (2001) 477] [arXiv:hep-th/9906070]. arXiv:hep-th/9702115,

\bibitem{WB} J. Wess, J. Bagger: Supersymmetry and Supergravity, Princeton
University Press, Princeton, 1983.
\end{thebibliography}
\end{document}